\documentclass[sigconf]{acmart}
\AtBeginDocument{%
	}

\copyrightyear{2026}
\acmYear{2026}
\setcopyright{cc}
\setcctype{by}
\acmConference[E-Energy '26]{The 17th ACM International Conference on Future and Sustainable Energy Systems}{June 22--25, 2026}{Banff, AB, Canada}
\acmBooktitle{The 17th ACM International Conference on Future and Sustainable Energy Systems (E-Energy '26), June 22--25, 2026, Banff, AB, Canada}
\acmDOI{10.1145/3744255.3811726}
\acmISBN{979-8-4007-2011-6/2026/06}

\usepackage[]{hyperref}
\usepackage{graphicx} % DO NOT CHANGE THIS
\usepackage{subfigure}
\usepackage{booktabs}
\usepackage{xcolor}
\usepackage{xspace}
\usepackage{makecell}

\newcommand{\ouralg}{$\mathsf{WWS}$\xspace}
\newcommand{\carbon}{$\mathsf{CARBON}$\xspace}
\newcommand{\water}{$\mathsf{WATER}$\xspace}
\newcommand{\eWater}{$\mathsf{STRESS}$\xspace}

\begin{document}

\title{Balancing Bits and Drops: Stress-Adjusted Water Management for Data Centers}

\author{Zahidur Talukder}
\authornote{Part of this work was done at the University of Texas at Arlington.}
\affiliation{%
	\department{Department of Math and Computer Science}
	\institution{Texas Lutheran University}
	\city{Seguin}
	\state{Texas}
	\country{USA}
}
\email{ztalukder@tlu.edu}

\author{Imtiaz Bin Rahim}
\affiliation{%
	\department{Department of Computer Science and Engineering}
	\institution{University of Texas at Arlington}
	\city{Arlington}
	\state{Texas}
	\country{USA}
}
\email{ixb6394@mavs.uta.edu}

\author{Pranjol Sen Gupta}
\affiliation{%
	\department{Department of Computer Science}
	\institution{Kennesaw State University}
	\city{Kennesaw}
	\state{Georgia}
	\country{USA}
}
\email{pgupta10@kennesaw.edu}

\author{Shaolei Ren}

\affiliation{%
	\department{Department of Electrical and Computer Engineering}
	\institution{University of California, Riverside}
	\city{Riverside}
	\state{California}
	\country{USA}
}
\email{shaolei@ucr.edu}

\author{Mohammad A. Islam}
\affiliation{%
	\department{Department of Computer Science and Engineering}
	\institution{University of Texas at Arlington}
	\city{Arlington}
	\state{Texas}
	\country{USA}
}
\email{mislam@uta.edu}

\begin{abstract}
Data centers are critical to today’s digital economy, but are also among the largest industrial consumers of freshwater. 
Beyond the sheer volume of water use, the environmental impact of data center water consumption varies significantly across locations and seasons, depending on local and regional water stress.
However, prior research has largely focused on reducing total water use, overlooking that the same unit of water can have drastically different environmental consequences depending on when and where it is consumed. In this paper, we introduce a \emph{stress-adjusted water} framework that quantifies the true sustainability impact of data center water consumption by incorporating both spatial and temporal water stress.
Using the AWARE-US model, we capture county-level monthly variations in water availability and extend this framework to account for the off-site water footprint of electricity generation.
Based on this stress-aware accounting, we analyze stress-adjusted water-computing strategies spanning both the software and infrastructure layers.
Specifically, we study workload scheduling policies that jointly optimize water and carbon efficiency, evaluate the potential of rainwater harvesting as a supplemental water source, and investigate the feasibility of dry cooling as a water-free alternative to evaporative cooling. Our evaluation across major U.S. data center markets shows that stress-aware workload scheduling can reduce stress-adjusted water consumption by up to 25\% while preserving performance and balancing carbon emissions.
We further show that rainwater harvesting can offset up to 100\% of on-site cooling water in regions with sufficient precipitation, while offering diminishing returns in arid locations.
Finally, our analysis of dry cooling reveals that its effectiveness depends critically on energy efficiency and off-site water stress, highlighting important trade-offs between water savings, energy use, and carbon emissions.
Together, these results demonstrate that incorporating water stress as a first-class sustainability metric enables more informed and effective data center design and operation.
\end{abstract}

\begin{CCSXML}
	<ccs2012>
	<concept>
	<concept_id>10010583.10010662.10010673</concept_id>
	<concept_desc>Hardware~Impact on the environment</concept_desc>
	<concept_significance>500</concept_significance>
	</concept>
	<concept>
	<concept_id>10002944.10011123.10010912</concept_id>
	<concept_desc>General and reference~Empirical studies</concept_desc>
	<concept_significance>300</concept_significance>
	</concept>
	</ccs2012>
\end{CCSXML}

\ccsdesc[500]{Hardware~Impact on the environment}
\ccsdesc[300]{General and reference~Empirical studies}

\keywords{Data centers, water sustainability, water stress, workload scheduling, rainwater harvesting, dry cooling}

\maketitle % should come after the abstract

\section{Introduction}

%\textbf{Importance of water.}
Water is one of the most abundant natural resources on Earth, yet access to freshwater remains limited and unevenly distributed. The global water crisis is further intensified by climate change, rapid population growth, and aging infrastructure. The United Nations (UN) highlights water scarcity as one of the most pressing consequences of climate change, underscoring the urgency of global action \cite{UN2020}. Since water is a shared societal resource, industries across all sectors must actively contribute to sustainability efforts. Among them, data centers—given their immense scale and critical role in the digital economy—have both the opportunity and the responsibility to lead by example in advancing sustainable water management \cite{Mekonnen2015}.

\textbf{Water consumption in data centers.}
Data centers are widely recognized for their massive energy consumption, but they are also huge water guzzlers \cite{gnibga2024flexcooldc,li2023making}.
Due to their high server cooling loads, most large-scale data centers rely on highly efficient cooling tower systems that use water evaporation to expel heat. However, this evaporation leads to continuous water loss, requiring constant replenishment \cite{shehabi20242024}.
In 2023 alone, Google’s self-operated data centers withdrew approximately 29 billion liters of water for on-site cooling, with over 23 billion liters evaporated—nearly 80\% of which was potable water \cite{GoogleSustainability2023}. As the demand for digital services surges, driven largely by advances in artificial intelligence (AI) and machine learning (ML), data center water usage is rising at an unprecedented rate. From 2021 to 2022, Google’s water consumption increased by around 20\%, followed by another 17\% increase from 2022 to 2023. Similarly, Microsoft reported a 34\% increase between 2021 and 2022, and a further 22\% rise from 2022 to 2023 \cite{MicrosoftESG2023}.
% \note{check 2024 data}.
The U.S. Department of Energy projects that, by 2028, total annual on-site water consumption by U.S. data centers could double or even quadruple from 2023 levels \cite{USDOE2023}.
Beyond direct water use for cooling, data centers are also responsible for the water consumption associated with their grid electricity use \cite{islam2016exploiting,siddik2021environmental}. 
Electricity generation from nuclear and thermal energy sources also requires water in their cooling systems. 
Excluding hydroelectricity, which itself is a major water consumer, the national average water consumption by electricity power plants in the U.S. is as high as 1.8 L/kWh \cite{Grubert2018}.

\textbf{Limitations of prior works.}
Over the past decade, data centers have made substantial progress in improving energy efficiency and reducing carbon emissions through carbon-aware scheduling, placement, and infrastructure design \cite{gupta2021chasing,li2023toward,sukprasert2024limitations}.
In contrast, data center water sustainability has largely been studied through the lens of \emph{total water consumption}, often mirroring carbon-centric optimization approaches \cite{islam2014exploiting,islam2015water,islam2016exploiting}.
While these efforts provide valuable insights into improving water efficiency, they overlook a fundamental distinction between the environmental impacts of water consumption and carbon emissions.
Carbon emissions contribute to climate change at a global scale, and their marginal impact is largely independent of where or when they occur \cite{IPCC2021}.
In contrast, the environmental impact of water consumption is inherently \emph{local and time dependent} \cite{lee2019aware,lee2020regional,Mekonnen2016}.
Consuming one liter of water in a water-stressed region during a dry season can have orders-of-magnitude greater ecological and societal consequences than consuming the same amount of water in a water-abundant region or during a wet season.
As a result, treating all water consumption as environmentally equivalent fundamentally misrepresents its true impact.

Recent work has begun to move beyond purely volumetric accounting by explicitly valuing water based on its scarcity.
In a recent presentation, the Open Compute Project (OCP) introduced an updated water-efficiency metric that adjusts for regional water scarcity, expressed as the ratio of water withdrawal to water availability \cite{vardhan2025water}.
Wu et al.\ proposed SCARF, which introduces an Adjusted Water Impact (AWI) metric to reflect regional water scarcity when evaluating computing workloads \cite{wu2025not}.
SCARF demonstrates that treating water as location-invariant can substantially alter sustainability conclusions. However, it applies regional adjustment at an aggregate level and does not explicitly distinguish between \emph{on-site} and \emph{off-site} water pathways or model the provenance of electricity generation.
More closely related in spirit, Jiang et al.\ proposed \emph{WaterWise}, a scheduler that incorporates water stress into workload scheduling and shows that optimizing for water and carbon independently can yield conflicting outcomes \cite{jiang2025waterwise}.
WaterWise models off-site water impact using region-level abstractions and does not explicitly account for the geographic locations of water-consuming power plants, which can misestimate the water stress borne by electricity generation—especially when data centers draw power from stressed generation sites located far from the data center itself.

\textbf{Our key insight and contributions.}
This paper builds on and extends these efforts by explicitly modeling the \emph{provenance} of both on-site and off-site water consumption.
Our key insight is that sustainable data center operation must reason about the \emph{value of water}\footnote{We interpret the value of water based on availability; its value rises with scarcity and falls with abundance.}, not merely its volume.
Specifically, the environmental cost of water consumption depends on \emph{where} and \emph{when} the water is consumed, as well as whether it occurs directly at the data center or indirectly through electricity generation.

To capture these effects, we leverage the \textbf{A}vailable \textbf{WA}ter \textbf{RE}\allowbreak maining for the \textbf{U}nited \textbf{S}tates (AWARE-US) model, which provides county-level monthly water-stress characterization across the U.S.~\cite{lee2019aware}.
Using AWARE-US, we introduce the notion of \emph{stress-adjusted water consumption}, which weights water usage by local water scarcity to reflect its true environmental impact.
We apply this metric to both direct (on-site) water consumption and indirect (off-site) water consumption embedded in electricity use.

While AWARE-US provides fine-grained characterization of \emph{local} water stress, it does not capture the off-site water impacts embedded in electricity generation.
To address this limitation, we explicitly trace data center electricity consumption to upstream power plants within the U.S. eGRID sub-regions \cite{epaEgrid} and estimate off-site water stress by weighting plant-level water stress according to electricity generation.
This enables a more accurate estimation of off-site water stress than region-level or national-average abstractions.
In practice, water intensity varies substantially across locations even for the same generation technology.
As shown in Table~\ref{tab:fixed_wue_error}, relying on fixed national water-intensity factors can lead to substantial misestimation of off-site water use.
Because stress-adjusted water combines both volumetric consumption and local water stress, such inaccuracies directly propagate into errors in the overall water-impact assessment.
Notably, minimizing volumetric water can misrank decisions under geographic flexibility and may even increase stress-adjusted water by shifting demand toward a more water-stressed electricity supply.

Using our framework, we analyze more than 4{,}000 U.S.\ data center locations obtained from \texttt{datacentermap.com}~\cite{DataCenterMapUSA}.
Our analysis reveals that nearly 75\% of U.S.\ data centers experience at least one month of elevated water stress annually, and that off-site water stress can dominate total water impact in many major markets.

\textbf{Stress-adjusted water computing strategies.}
Building on this framework, we investigate three complementary approaches for reducing stress-adjusted water impact in data center operations, spanning software-level optimizations and infrastructure-level interventions.

\textit{Workload scheduling.}
We study how spatial and temporal workload flexibility can be leveraged to reduce stress-adjusted water consumption by preferentially executing workloads at locations and times with lower water stress.
Our analysis considers multiple workload classes with varying degrees of flexibility and jointly optimizes for water and carbon objectives.
We show that stress-aware scheduling can reduce stress-adjusted water consumption by up to 25\% while simultaneously reducing carbon emissions by up to 25\%.

\textit{Rainwater harvesting.}
We evaluate rainwater harvesting as a location-dependent mitigation strategy that reduces reliance on stressed freshwater sources.
Using year-long precipitation data across major data center markets, we analyze the feasibility of rainwater harvesting under realistic storage constraints.
Our results demonstrate that, in suitable regions, rainwater harvesting can offset up to 100\% of on-site cooling water demand.

\textit{Dry cooling.}
Finally, we examine the water-energy trade-offs introduced by dry-cooling technologies, which eliminate on-site water use but incur higher electricity consumption \cite{shehabi20242024}.
While dry cooling reduces direct water withdrawal, the resulting increase in electricity demand shifts water consumption off-site to power plants, where water stress may be significantly higher.
We quantify this trade-off across major U.S. data center markets and show that dry cooling is beneficial primarily in highly water-stressed regions when the associated energy efficiency penalty remains sufficiently low.

\textbf{Dataset.}
We construct a dataset comprising 4{,}147 U.S.\ data center locations annotated with on-site and off-site water stress metrics, electricity provenance, regional precipitation profiles, cooling efficiencies, and electricity-generation water efficiencies.
To support transparency and future research, our dataset is made publicly available on the Open Science Framework (OSF) \cite{osf_water_stress_dataset}.

\textbf{Limitations.}
Our study has several limitations that point to future research directions.
We rely on AWARE-US for water stress characterization; alternative hydrological models may yield different absolute stress values.
Our off-site analysis uses publicly available EPA power plant data, which may omit small or recently commissioned plants.
Rainwater harvesting is evaluated using precipitation data from a single year (2023), and inter-annual climate variability may affect long-term outcomes.
Our workload scheduling analysis provides an offline upper bound rather than an online scheduling algorithm.
Finally, data center location data represents a snapshot in time and may not capture newly deployed facilities.

Looking forward, we believe that combining stress-aware water accounting with demographic and equity analyses—such as assessing the co-location of data centers with vulnerable or underserved communities—represents an important and largely unexplored direction for sustainable computing research.

\section{Preliminaries}

\subsection{Water Usage}

The data center community often discusses ``water usage'' without carefully distinguishing among \textit{withdrawal}, \textit{discharge}, and \textit{consumption}.
These terms are not interchangeable and lead to very different sustainability conclusions \cite{mytton2021data}.
\textit{Water withdrawal} refers to the total volume of water taken from a source (e.g., a municipal utility, river intake, or groundwater well).
\textit{Water discharge} is the portion returned to the environment after use (e.g., cooling-water blowdown or once-through return flow).
\textit{Water consumption} is the fraction of withdrawn water that is not returned to the original watershed, typically because it evaporates, is incorporated into products, or is transported to another watershed \cite{mytton2021data}.
Thus, a facility’s water consumption can be expressed as:
\begin{equation}
\text{consumption} = \text{withdrawal} - \text{discharge}.
\end{equation}

\begin{figure}
	\centering
	\includegraphics[page=1, trim = 6cm 5cm 0.5cm 5cm, clip, width=1\linewidth]{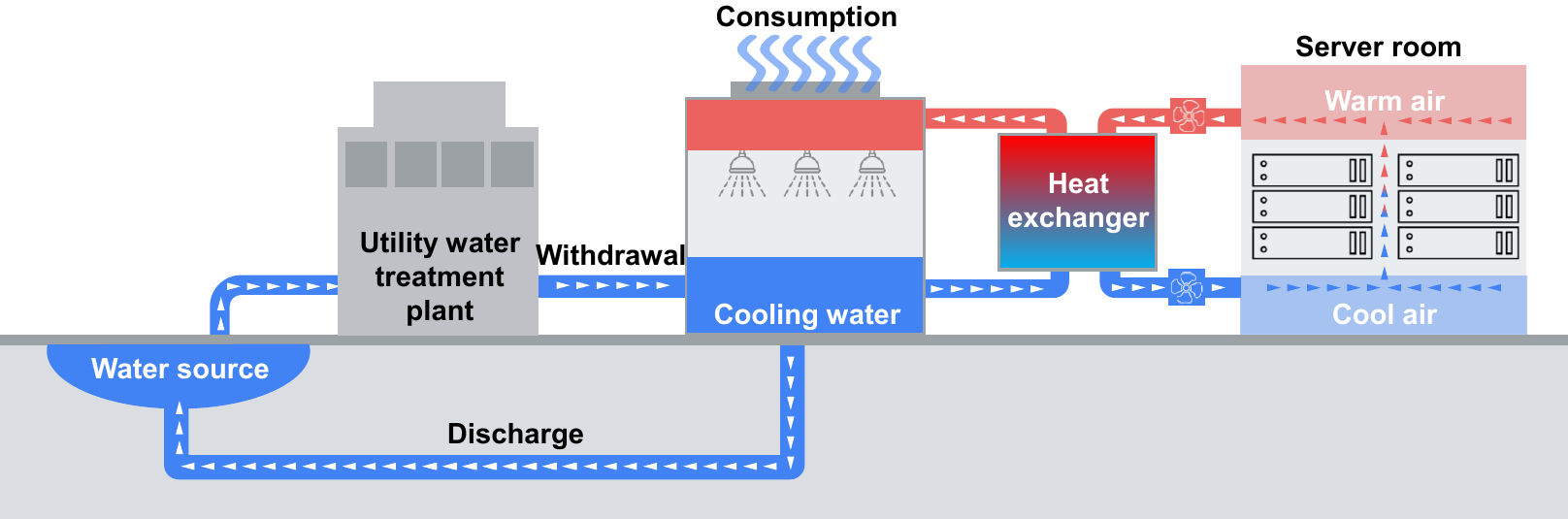}
	\caption{Water usage in data center heat rejection.}
	\label{fig:water_cooling}
\end{figure}

Fig.~\ref{fig:water_cooling} illustrates these concepts for cooling-tower-based data center operation.
In evaporative cooling, a portion of circulating water evaporates to reject heat, which directly contributes to \textit{consumption}.
Another portion is periodically discharged (``blowdown'') to control mineral concentration, which contributes to \textit{discharge}.
Cooling systems typically cannot achieve both low withdrawal and zero consumption simultaneously.
For example, \textit{once-through} cooling can achieve near-zero consumption by returning most of the withdrawn water, but it requires far larger withdrawal volumes to remove the same heat load \cite{lee2020regional}.
These trade-offs matter because they affect different stakeholders: withdrawal stresses water \textit{supply capacity} and infrastructure (treatment, pumping, distribution), while consumption reduces \textit{availability} for competing societal and ecological demands.

Water usage is often categorized by color to distinguish its source, quality, and suitability for various uses. 
\textit{Blue Water} refers to surface and groundwater (e.g., lakes, rivers, and aquifers) that are withdrawn for human use, such as cooling data centers and industrial processes. Once consumed (e.g., through evaporation or incorporation into products), it is no longer available in the original source \cite{hoekstra2012water}. \textit{Green Water} represents rainwater and soil moisture that is naturally stored in the unsaturated zone of soil and used by plants for growth. It is crucial for agriculture and forestry, but is generally not part of direct industrial or urban water consumption \cite{falkenmark2004balancing}. \textit{Grey Water} is wastewater from domestic, commercial, or industrial activities that has been used but is not heavily contaminated (e.g., from sinks, showers, or cooling processes). It can be treated and reused for non-potable purposes such as irrigation or industrial cooling \cite{zhuo2016inter}.

\subsection{Water Consumption in Data Centers}
To systematically reason about water impacts, we adapt the well-known sustainability scope framework from the Greenhouse Gas Protocol \cite{GHGProtocol2024}.
Although originally designed for carbon accounting, the scope decomposition is also useful for water because it separates operational impacts from supply-chain impacts and clarifies which levers are controllable during operation.

\textbf{Scope 1 (direct on-site water consumption).}
This includes water consumed within the data center boundary, primarily for cooling (e.g., evaporation in cooling towers), humidification, and other facility processes.
In Fig.~\ref{fig:water_cooling}, evaporated cooling water constitutes the dominant Scope~1 component in many large-scale facilities.

\textbf{Scope 2 (indirect off-site water consumption from electricity).}
This includes water consumed upstream by electricity generation needed to power IT and cooling loads.
Thermal and nuclear power plants often consume water through evaporative cooling, and this consumption occurs at the plants rather than at the data center.
As a result, the off-site water footprint can be geographically decoupled from the facility and may be subject to very different water scarcity conditions.

\textbf{Scope 3 (other indirect water consumption).}
This includes embodied water from manufacturing IT equipment, constructing facilities, and broader supply chain processes.
While Scope~3 can be significant, it is difficult to track at high resolution and harder to mitigate through operational decisions.
Accordingly, we focus on \textbf{operational water consumption} (Scopes~1 and~2) that can be influenced by water-aware computing and infrastructure choices.

\subsection{Water Usage Effectiveness (WUE)}
WUE is a widely used metric proposed by The Green Grid to quantify water efficiency \cite{azevedo2011water}.
For on-site operations, WUE captures facility water consumption per unit of IT energy (L/kWh), making it suitable for comparing cooling and facility designs.
For off-site electricity, the analogous concept is sometimes referred to as the Electricity Water Intensity Factor (EWIF) \cite{azevedo2011water,islam2015water}.
Because grid electricity comes from a portfolio of generation sources with different water intensities, off-site WUE is often computed as a generation-weighted average:
\begin{align}\label{eq:wue_Offsite}
WUE_{offsite} = \frac{\sum_i G_i \cdot WUE_i}{\sum_i G_i},
\end{align}
where $G_i$ is the electricity generated by source $i$ and $WUE_i$ is its water intensity.
This formulation is attractive because hourly generation data by source is often publicly available \cite{gupta2024dataset}.
However, the approach is only as accurate as the assumed $WUE_i$ values.
A common simplification is to treat fuel-specific WUE values as fixed across large geographies (e.g., a single U.S.-wide WUE for natural gas), which, as shown in Table~\ref{tab:fixed_wue_error}, can introduce regional errors.

OCP has recently introduced an updated WUE metric, WUE$^+$, that adds two multiplicative factors to the original WUE to capture regional water availability and reuse efficiency \cite{vardhan2025water}. It is defined as follows
\begin{align}
	WUE^+ = WUE \times \text{Regional Adjustment} \times \text{Reuse Efficiency},
\end{align} 
where regional adjustment is defined as the ratio of total water withdrawal and available renewable water, and reuse efficiency is defined as the ratio of potable water used and total water input.

\subsection{Variation in WUE}
On-site and off-site water efficiency varies across geography and time.
On-site WUE depends on local weather (temperature and humidity), cooling tower characteristics, and facility design \cite{gupta2024dataset}.
For the same IT load, warmer or more humid conditions can lead to higher evaporation and, in turn, higher on-site water consumption.
Off-site WUE depends on the regional generation mix, power plant cooling technologies, and local operating conditions.

\begin{figure}
	\centering
	\subfigure[On-site WUE]{\includegraphics[width=0.48\linewidth]{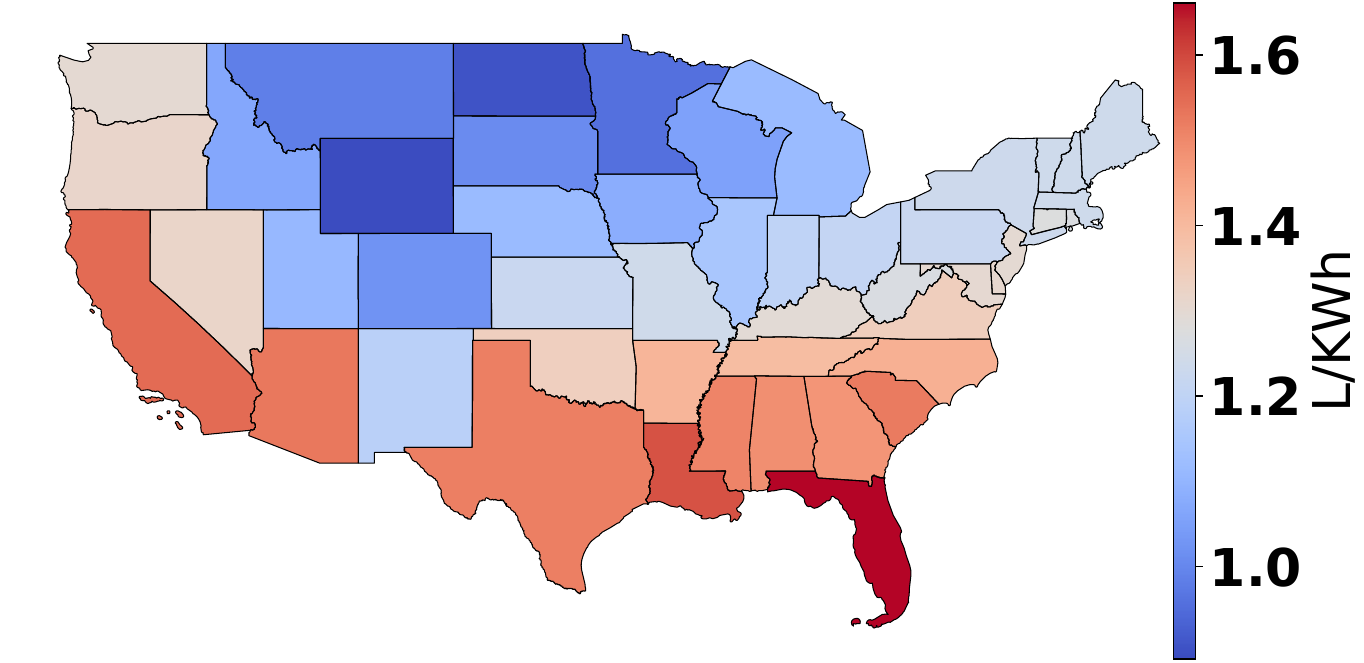}} 
	\subfigure[Off-site WUE]{\includegraphics[width=0.48\linewidth]{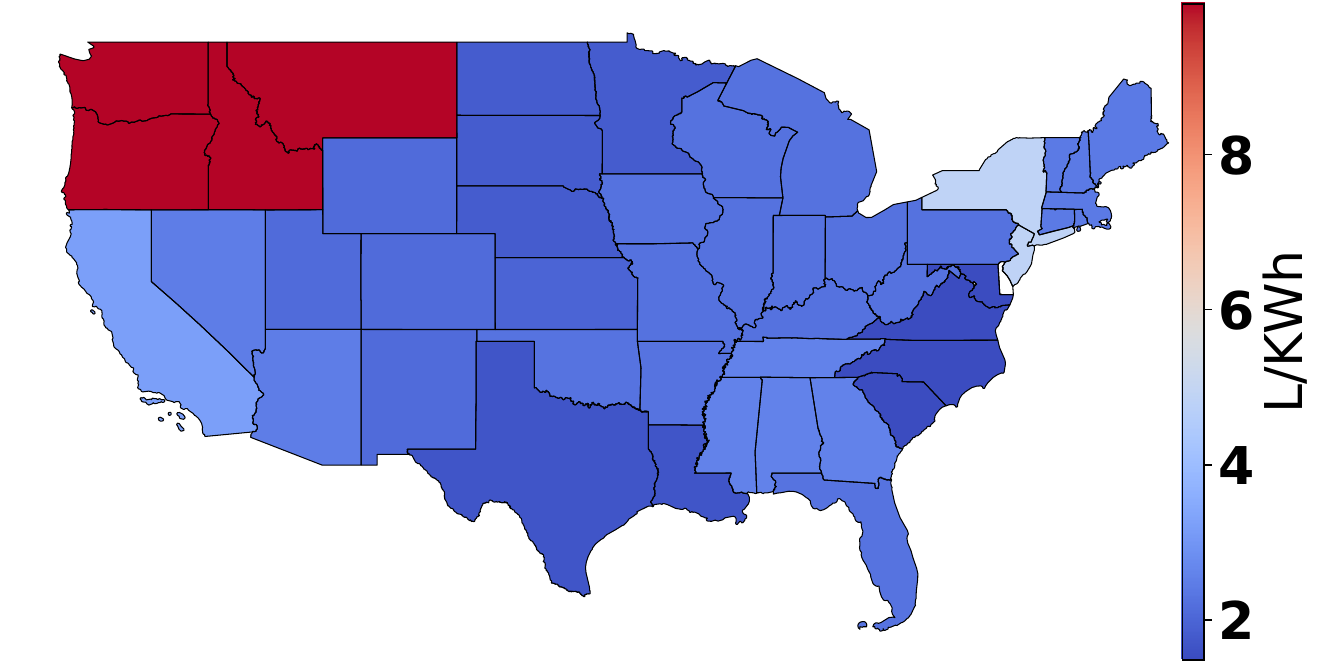}}
%	\subfigure[Offsite carbon]{\includegraphics[width=0.32\linewidth, trim={1cm 3cm 0cm 3cm}, clip]{Figure/offsite_carbon_map.png}}
	\caption{Spatial variation in on-site and off-site water efficiency across U.S. states.}
	\label{fig:variation_spatial_WUE}
\end{figure}

Fig.~\ref{fig:variation_spatial_WUE} illustrates spatial variation in average on-site and off-site WUE across the U.S.
We observe substantial regional diversity, with some states exhibiting markedly higher off-site WUE due to reliance on water-intensive generation technologies.
This heterogeneity motivates software strategies (e.g., workload shifting) that exploit spatiotemporal differences in WUE.

% \textbf{Efficiency is not impact.}
While WUE is useful, it measures \emph{liters per kWh}, not the environmental cost of consuming those liters.
Two regions can have the same WUE but very different water scarcity conditions, meaning equal volumes of water consumption may have dramatically different societal and ecological consequences.
This motivates stress-aware accounting, which we introduce next.

\section{Stress-Adjusted Water Footprint}
\label{sec:stress_adjusted_water_footprint}

\subsection{Water versus Carbon}
A major reason water has been historically underemphasized in computing sustainability is the community’s success with carbon-aware optimization.
Carbon emissions are an appropriate global sustainability metric because greenhouse gases mix in the atmosphere and persist over long time horizons; therefore, marginal damages are often treated as largely independent of where and when emissions occur \cite{IPCC2021}.
This justifies objectives such as minimizing total operational emissions or shifting workloads to regions with low-carbon electricity.

Water consumption is fundamentally different.
Water is locally sourced, locally constrained, and shared among households, agriculture, ecosystems, and industry.
Thus, the environmental impact of consuming one liter of freshwater depends strongly on local scarcity \cite{lee2019aware,lee2020regional,Mekonnen2016}.
Moreover, water availability is time-varying: seasonal patterns and drought conditions can sharply alter scarcity and competition for water.
As a result, minimizing total water volume or maximizing WUE is insufficient to assess the sustainability impact.

\subsection{Water Stress}
Water stress refers to the pressure on local water resources resulting from the balance between availability and demand.
Many tools quantify water stress at different spatial and temporal scales \cite{lee2019aware,droughMonitor,wriAquaduct}.
We use the AWARE-US model because it provides county-level monthly characterization factors across the U.S. and has been adopted in prior assessments of electricity-related water impacts \cite{lee2019aware}.

AWARE-US defines a unit-less characterization factor (CF) that converts each liter of water consumption into a \emph{stress-adjusted} quantity.
CF is derived from availability-minus-demand (AMD), where AMD is computed from hydrological runoff and social/environmental water demands.
The CF of county $i$ in month $j$ is:
\begin{align}
CF_{i,j} = \frac{AMD_{US}}{AMD_{i,j}},
\end{align}
where $AMD_{US}$ is the U.S. reference value \cite{lee2019aware}.
CF is capped between 0.1 and 100 to avoid extreme values dominating assessments.
In our interpretation, $CF>1$ indicates above-average water stress.
\begin{figure}
\centering
\includegraphics[width=\linewidth]{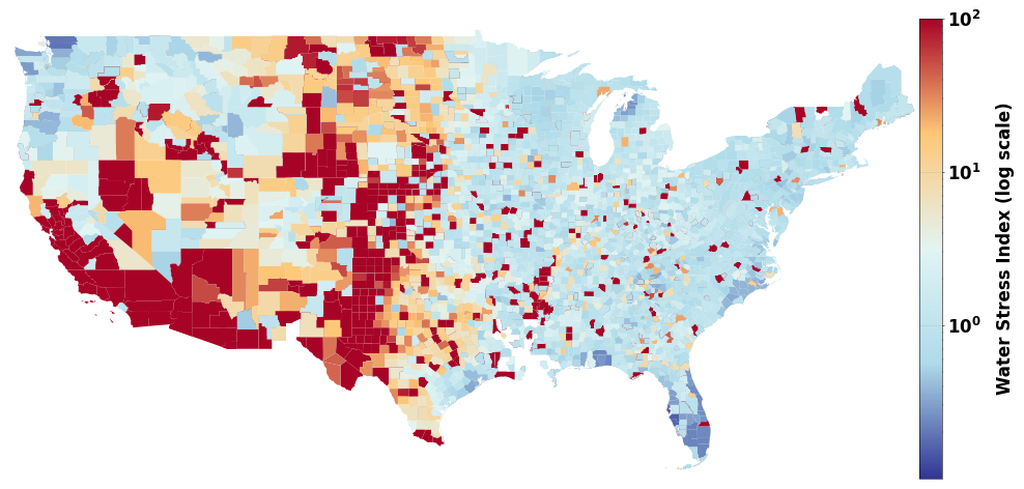}
\caption{County-level water stress across the United States based on the AWARE-US model, illustrating strong spatial heterogeneity in water stress \cite{lee2019aware}.}
\label{fig:stress_map}
\end{figure}

\begin{figure}
	\centering
	\includegraphics[width=\linewidth]{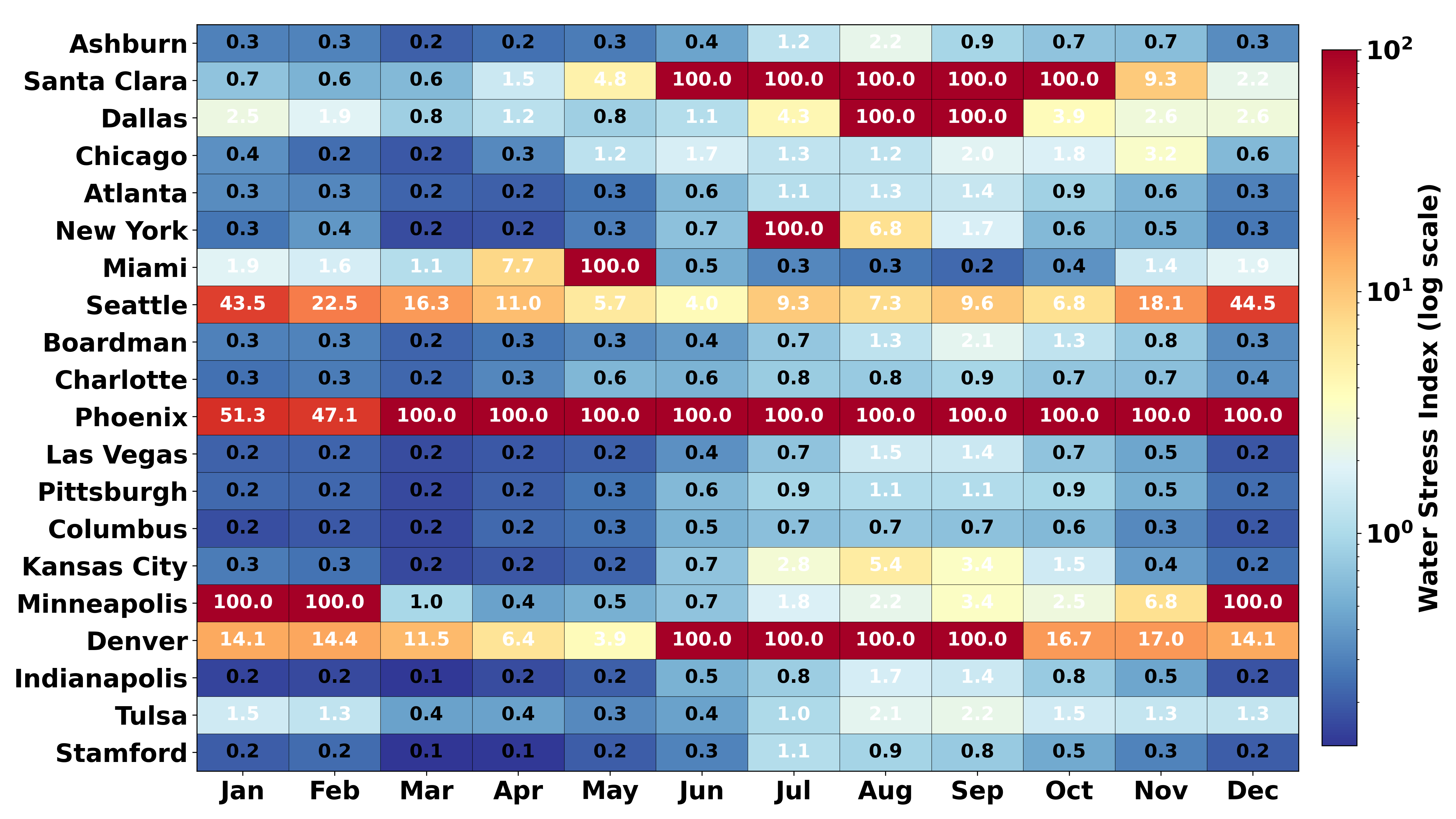}
	\caption{Monthly water stress across major U.S. data center markets, illustrating both spatial variation across locations and temporal variation throughout the year.}
	\label{fig:data_center_monthly_stress}
\end{figure}

\textbf{Spatiotemporal variation.}
Water stress exhibits pronounced heterogeneity across both geography and time due to seasonal hydrology and regional demand patterns. Fig.~\ref{fig:stress_map} illustrates county-level water stress during the summer, revealing strong spatial disparities in water availability across the United States at a single point in time. Fig.~\ref{fig:data_center_monthly_stress} further highlights this variability by showing monthly water stress across the top 20 U.S. data center markets, where stress levels fluctuate substantially throughout the year and differ markedly between locations. Together, these results demonstrate that water stress is inherently spatiotemporal, implying that the environmental impact or \emph{value} of a unit of water consumption can vary significantly depending on when and where it occurs. This variability motivates the need for stress-adjusted water accounting rather than volume-based metrics.

\subsection{Stress-Adjusted Water Consumption}

We define \textbf{stress-adjusted water} as water consumption weighted by local stress conditions.

\textbf{On-site stress-adjusted water.}
For a data center located in county $c$ and month $m$, on-site stress-adjusted water is computed as:
\begin{equation}
W^{on}_{SA}(m) = W^{on}(m)\cdot CF_{c,m},
\end{equation}
where $W^{on}(m)$ is the on-site water consumption in month $m$.

\textbf{Off-site stress-adjusted water.}
Off-site water consumption occurs at power plants due to electricity generation and is therefore governed by the water stress at plant locations rather than at the data center site.
Estimating off-site stress-adjusted water thus requires mapping a data center’s electricity consumption to the upstream generation portfolio supplying its electricity.

\begin{figure}[t!]
	\centering
	\includegraphics[width=1.0\linewidth]{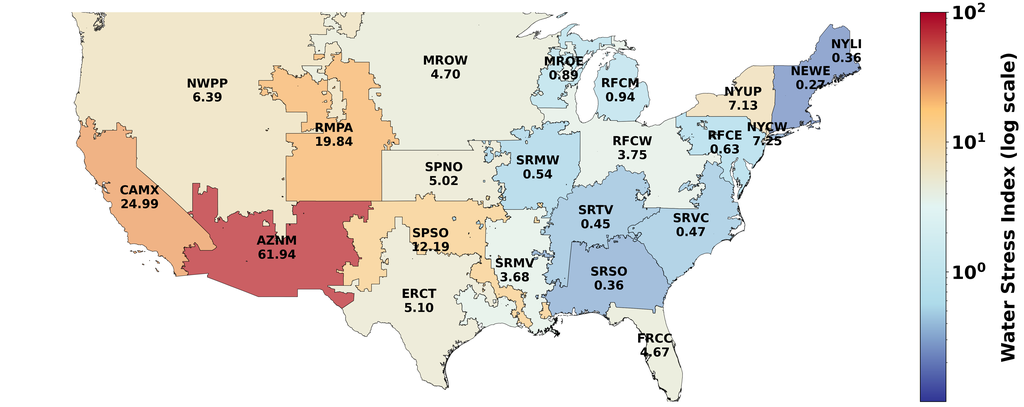}
	\caption{Off-site water stress across EPA eGRID sub-regions, computed using generation-weighted AWARE-US characterization factors.}
	\label{fig:egrid_poc_figure}
\end{figure}

\begin{table*}[!ht]
	\caption{Water efficiency in L/kWh of different fuel sources across the U.S. eGRID regions. ``-'' indicates no electricity generation from the corresponding source. ``0'' indicates no water consumption (e.g., once-through cooling) during electricity generation.}\label{tab:water_efficiecny}
	\centering
	\resizebox{\linewidth}{!}{%
		\begin{tabular}{|l|l|l|l|l|l|l|l|l|l|l|l|l|l|l|l|l|l|l|l|l|l|l|l|}
			\hline
			\textbf{Primary Fuel} & \textbf{AZNM} & \textbf{CAMX} & \textbf{ERCT} & \textbf{FRCC} & \textbf{HIOA} & \textbf{MROE} & \textbf{MROW} & \textbf{NEWE} & \textbf{NWPP} & \textbf{NYCW} & \textbf{NYLI} & \textbf{NYUP} & \textbf{RFCE} & \textbf{RFCM} & \textbf{RFCW} & \textbf{RMPA} & \textbf{SPNO} & \textbf{SPSO} & \textbf{SRMV} & \textbf{SRMW} & \textbf{SRSO} & \textbf{SRTV} & \textbf{SRVC} \\ \hline
			\textbf{COAL}         & 1.82          & 1.35          & 1.68          & 1.42          & 2.04          & \_            & 1.23          & 0             & 2.19          & \_            & \_            & \_            & 2.21          & 1.12          & 1.35          & 1.83          & 1.44          & 2.37          & 1.27          & 2.42          & 2.24          & 1.57          & 1.29          \\ \hline
			\textbf{NATURAL GAS}  & 2.47          & 1.25          & 2.4           & 2.58          & \_            & 0             & 1.63          & 1.01          & 1.74          & 0             & 0             & 0.69          & 1.71          & 2.21          & 1.56          & 1.46          & 2.13          & 3.23          & 1.86          & 1.78          & 2.48          & 1.38          & 1.67          \\ \hline
			\textbf{NUCLEAR}      & 2.83          & 0             & 0.88          & 0             & \_            & \_            & 1.95          & 0             & 2.67          & \_            & \_            & \_            & 1.29          & 3.14          & 1.33          & \_            & 2.16          & \_            & 1.86          & 2.01          & 2.72          & 0.21          & 0.52          \\ \hline
			\textbf{OTHER}        & 6.49          & 2.59          & 1.82          & 4.8           & \_            & 0             & \_            & 3.87          & 2.13          & \_            & \_            & \_            & 17.2          & 0             & 49.9          & 1.77          & \_            & 1.35          & 0.71          & \_            & 0.64          & 3.35          & 1.26          \\ \hline
			\textbf{PETROLEUM}    & \_            & \_            & \_            & \_            & 0             & \_            & \_            & 0.23          &               &               & \_            & 0             & 0.41          & \_            & 0             & \_            & \_            & 0             & 1.04          & \_            & \_            & \_            & 0             \\ \hline
			\textbf{SOLAR}        & \_            & 0             & \_            & \_            & \_            & \_            & \_            & \_            & \_            & \_            & \_            & \_            & \_            & \_            & \_            & \_            & \_            & \_            & \_            & \_            & \_            & \_            & \_            \\ \hline

		\end{tabular}%
	}
\end{table*}

\begin{table*}[t!]
	\centering
	\caption{Error introduced by using fixed national water-intensity factors instead of region-specific values.
		Min/Max are computed across eGRID sub-regions using Table~\ref{tab:water_efficiecny}.}
	\label{tab:fixed_wue_error}
	\resizebox{0.7\linewidth}{!}{%
		\begin{tabular}{lcccccc}
			\hline
			\textbf{Fuel} &
			\makecell{\textbf{Fixed}\\\textbf{WUE} \cite{reig2020guidance}} &
			\makecell{\textbf{Minimum}\\\textbf{Regional}} &
			\makecell{\textbf{Maximum}\\\textbf{Regional}} &
			\makecell{\textbf{Maximum}\\\textbf{Overestimation}} &
			\makecell{\textbf{Maximum}\\\textbf{Underestimation}} &
			\makecell{\textbf{Regional}\\\textbf{Variability}} \\
			& (L/kWh) & (L/kWh) & (L/kWh) & (\%) & (\%) & ($\times$) \\
			\hline
			Coal         & 1.82 & 1.12 & 2.42 & +63\%  & $-$25\% & 2.2$\times$ \\
			Natural Gas  & 0.80 & 0.69 & 3.23 & +15\%  & $-$75\% & 4.7$\times$ \\
			Nuclear      & 2.31 & 0.21 & 3.14 & +999\% & $-$26\% & 15.0$\times$ \\
			Petroleum    & 1.36 & 0.23 & 1.04 & +492\% & +31\%  & 4.5$\times$ \\
			Other        & 0.76 & 0.64 & 49.9 & +18\%  & $-$98\% & 78.0$\times$ \\
			\hline
	\end{tabular}}
\end{table*}

\subsection{Estimating Off-Site Water Stress}
Tracing the exact power flow from generators to a specific data center is challenging due to the physics of power networks, dispatch dynamics, and market operations.
Instead, we approximate electricity provenance using the EPA eGRID framework, which partitions the U.S. grid into 27 sub-regions \cite{epaEgrid,EPAeGRID2023}.
Within an eGRID sub-region, electricity is shared among consumers, and therefore, the regional generation portfolio provides a practical approximation for attributing off-site impacts.

We compute a region-level off-site stress factor by taking a generation-weighted average of county-level CF values of water-consuming power plants in that region:
\begin{align}
\overline{CF}_r = \frac{\sum_i G_i \cdot CF_i}{\sum_i G_i},
\end{align}
where $G_i$ is the annual net generation of plant $i$ and $CF_i$ is the CF of the county containing that plant.
We align monthly AWARE-US stress factors with monthly averaged generation profiles; annual generation shares are used only to approximate spatial provenance.
Fig.~\ref{fig:egrid_poc_figure} visualizes off-site stress across eGRID regions based on this metric.

\begin{figure}[!t]
	\centering
	\includegraphics[width=1\linewidth]{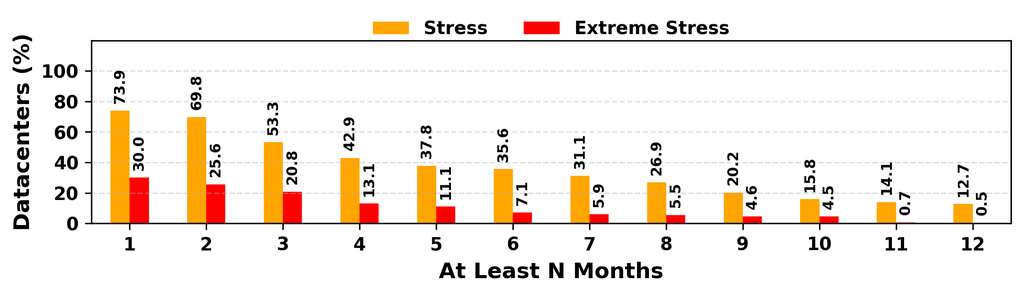}
	\caption{Percentage of U.S. data centers experiencing at least $N$ months of moderate or extreme on-site water stress per year.}
	\label{fig:data_center_stress}
\end{figure}

\subsection{U.S. Data Centers' Exposure to Water Stress}
Many major U.S. data center markets (e.g., California, Arizona, Texas) operate in regions experiencing substantial seasonal water scarcity.
To quantify the prevalence of water stress across the U.S. data center market, we analyze 4{,}147 data center locations from \texttt{datacentermap.com}.
Fig.~\ref{fig:data_center_stress} shows the distribution of data centers across stress levels. Here, we consider only the on-site water stress.
Nearly 75\% of data centers experience at least one month of water stress annually; approximately 30\% experience extreme stress (CF=100) in at least one month; and 12\% operate under persistent stress year-round.

\subsection{Limitation of On-Site Stress Alone}
\begin{figure}
	\centering
	\includegraphics[width=1\linewidth]{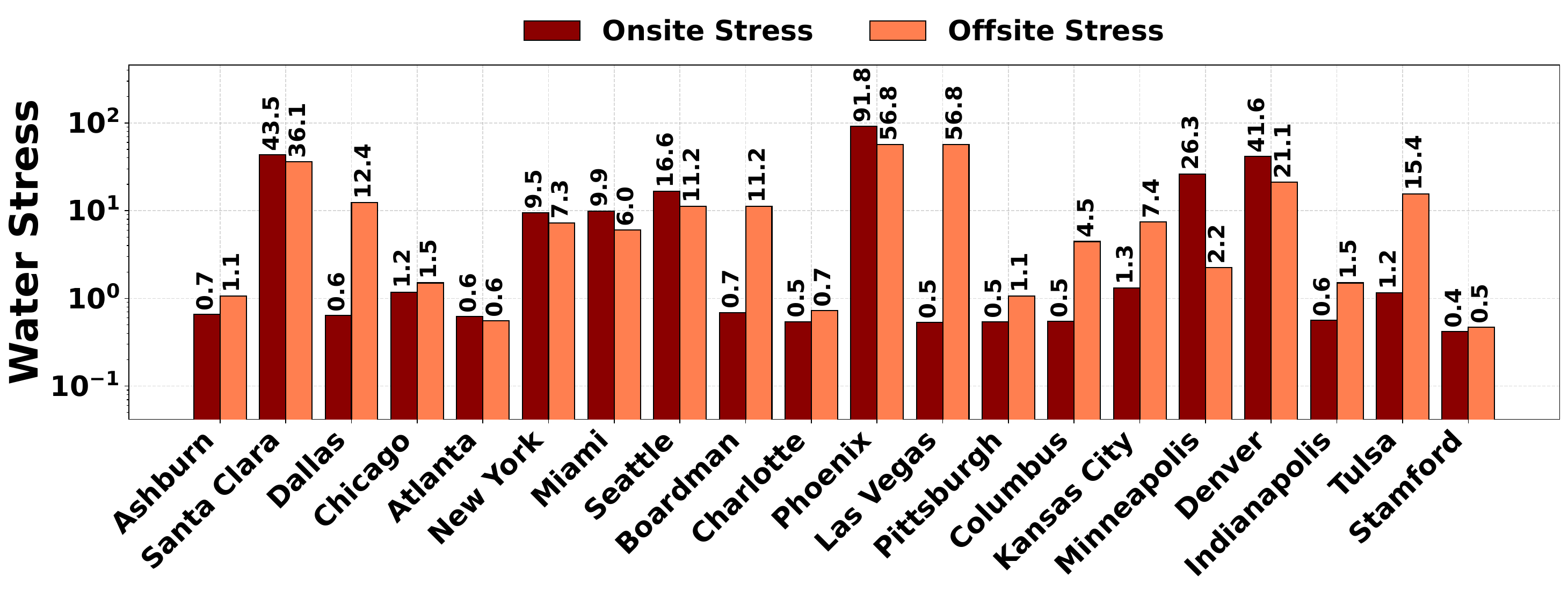}
	\caption{Comparison of on-site and off-site water stress across major U.S. data center markets.}
	\label{fig:average_water_stress_all_cities_barplot_log}
\end{figure}

While on-site CF captures local scarcity at the data center, the indirect footprint from electricity consumption can be influenced by very different water-stress conditions.
Fig.~\ref{fig:average_water_stress_all_cities_barplot_log} compares on-site and off-site water stress for the top U.S. data center markets.
We observe substantially higher off-site stress for several markets (e.g., Dallas, Boardman, Las Vegas, Columbus, Tulsa), indicating that a purely local assessment may underestimate the true stress burden of data center operation.
Conversely, some markets exhibit lower off-site stress, suggesting potential opportunities for water-aware workload distribution.

\subsection{Regional Off-Site Water Efficiency}
A second source of error in off-site water accounting arises from assuming fixed water intensities for generation sources across large regions.
In practice, even within the same fuel category (e.g., natural gas), water consumption per kWh varies across regions due to differences in cooling technology, plant design, local conditions, and operational practices.
To more accurately estimate off-site water consumption, we derive region-specific WUE values for each eGRID sub-region using 2023 power-plant-level water consumption data \cite{eiaWater}.
Table~\ref{tab:water_efficiecny} summarizes the resulting water efficiencies by generation source and region.

% \textbf{\textit{Key takeaway:}} Two data centers with identical WUE and total water consumption can differ by orders of magnitude in environmental impact once spatiotemporal water stress and electricity provenance are considered. Treating water as a homogeneous resource systematically misrepresents sustainability outcomes.

Table~\ref{tab:fixed_wue_error} quantifies the error introduced by using fixed national water-intensity factors, as is common in prior work. Across all fuel types, regional water intensity varies substantially, leading to systematic misestimation of off-site water consumption. For example, natural gas generation exhibits a 4.7× regional spread, leading to fixed values that underestimate water use by up to 75\% in some regions. For nuclear and “other” fuels, the error exceeds an order of magnitude. These results motivate our use of region-specific water efficiencies when estimating off-site, stress-adjusted water consumption.

\section{Stress-Adjusted Water Management}
\label{sec:water_wise}

Having established \emph{stress-adjusted water} as a metric that captures the \emph{value} of water consumption---accounting for where and when water is consumed and explicitly separating on-site and off-site impacts---we now examine how data centers can reduce their stress-adjusted water footprint in practice.
Rather than proposing a single mechanism, we analyze three complementary approaches that operate at different layers of the data center stack and involve distinct implementation costs and time horizons:
(i) \textbf{software-driven workload scheduling} that shifts computation across time and location,
(ii) \textbf{rainwater harvesting} that supplements on-site cooling demand with locally available precipitation,
and (iii) \textbf{dry cooling} that eliminates on-site water use by replacing evaporative cooling with air-based heat rejection.

Crucially, these approaches affect different components of the data center footprint.
Workload scheduling influences both \emph{on-site} (Scope~1) and \emph{off-site} (Scope~2) water consumption by changing when and where electricity is consumed, and cooling is performed.
Rainwater harvesting directly offsets \emph{on-site} (Scope~1) withdrawals without changing electricity demand.
In contrast, dry cooling eliminates on-site water use but increases electricity consumption, thereby shifting water impact from \emph{on-site} (Scope~1) to \emph{off-site} (Scope~2) through upstream power generation. Table~\ref{tab:cross_lever_summary} summarizes the overall contribution of three different techniques on stress-adjusted water and carbon.

\begin{table}[t]
	\centering
	\caption{Qualitative impact of \emph{stress-adjusted water} techniques on stress-adjusted water and carbon. Arrows indicate direction and relative magnitude.}
	\label{tab:cross_lever_summary}
	\resizebox{\columnwidth}{!}{%
		\begin{tabular}{lccc}
			\toprule
			\textbf{Technique} & \textbf{On-site Water} & \textbf{Off-site Water} & \textbf{Carbon} \\
			\midrule
			Workload Scheduling   & $\downarrow$ & $\downarrow$ & $\downarrow$ \\
			Rainwater Harvesting  & $\downarrow\downarrow$ & -- & -- \\
			Dry Cooling           & $\downarrow\downarrow$ & $\uparrow$ & $\uparrow$ \\
			\bottomrule
	\end{tabular}}
	
	\vspace{0.5em}
	\footnotesize
	$\downarrow\downarrow$: strong reduction,\;
	$\downarrow$: moderate reduction,\;
	$\uparrow$: moderate increase,\;
	--: negligible change.
\end{table}

\subsection{Workload Scheduling}
\label{sec:scheduling}

\subsubsection{Motivation: exploiting spatiotemporal heterogeneity}
On-site and off-site water impacts vary significantly across locations and over time due to weather-driven cooling dynamics, differences in power-grid composition, and geographic heterogeneity in water stress.
This spatiotemporal heterogeneity creates opportunities to reduce stress-adjusted water by shifting workloads toward \emph{low-impact} time--location pairs.
Modern cloud platforms already offer non-trivial scheduling flexibility, and prior work commonly categorizes workloads by their spatial and temporal freedom \cite{sukprasert2024limitations}.
We adopt three representative classes:

\textbf{Spatially flexible workloads.}
These workloads must be served immediately (tight response-time SLOs) but may be routed to one of several geographically distributed data centers.
Examples include web services, interactive applications, and ML inference.
Execution times typically range from milliseconds to seconds.

\textbf{Temporally flexible workloads.}
These workloads are delay-tolerant: they have a completion deadline (e.g., hours) and can be shifted in time within that window.
Examples include batch analytics, offline log processing, checkpointable HPC jobs, and many ML training pipelines.

\textbf{Spatio-temporally flexible workloads.}
These workloads can be shifted both in time and across locations, e.g., distributed training or migratable HPC jobs that can restart at alternative sites.
This class offers the largest optimization potential because it can exploit both temporal and geographic heterogeneity.

\begin{figure}[t!]
	\centering
	\includegraphics[width=\linewidth, trim= 0cm 0cm 0cm 3cm]{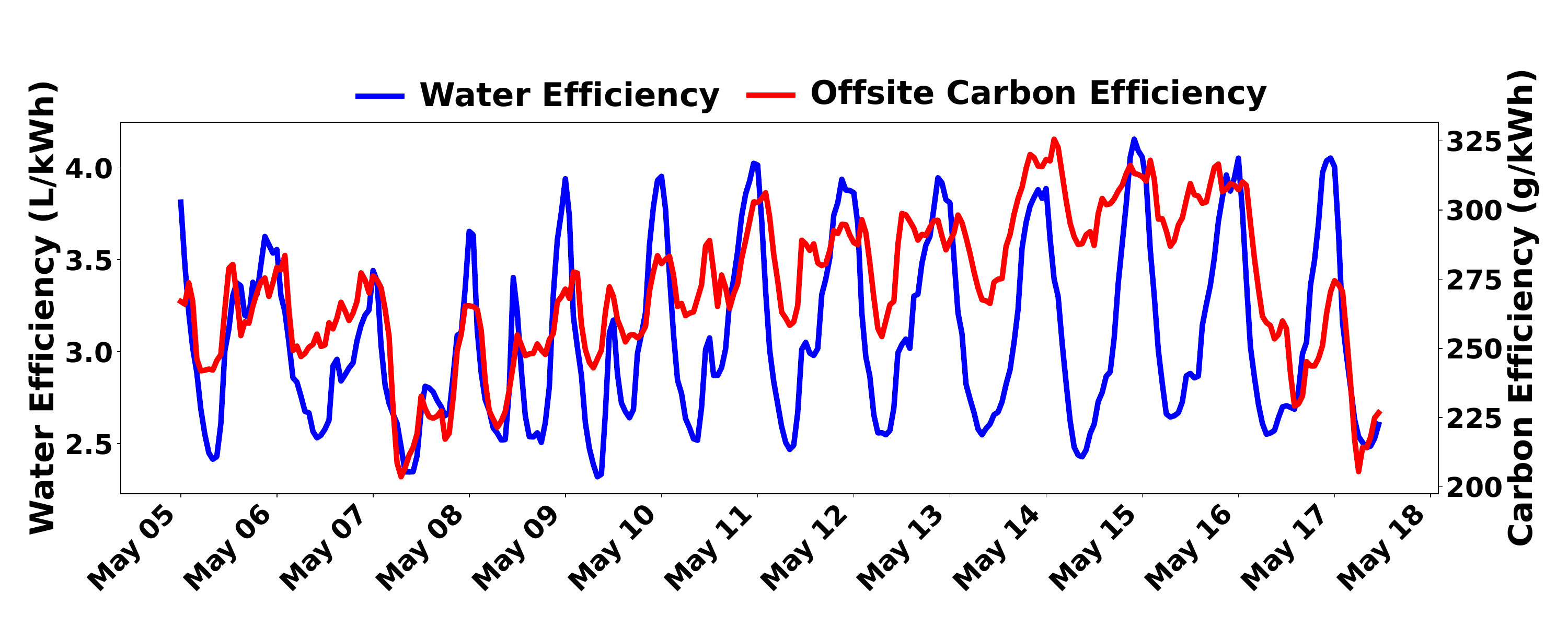}
	\caption{Hourly variation in water and carbon efficiencies (Ashburn, VA), illustrating weak correlation.}
	\label{fig:water_carbon_variation}
\end{figure}

\subsubsection{Scheduling overhead and performance assurance}
Scheduling flexibility is not free.
For spatial routing, sending an interactive request to a distant site increases network round-trip latency; satisfying the same SLO may therefore require additional provisioning or higher power.
For temporal or migratory jobs, overheads may arise from data transfer, checkpointing, and restart costs.
In our evaluation, we conservatively assume that \textbf{no performance degradation is permitted for sustainability}: any routing or deferral overhead is absorbed as additional power, ensuring SLO compliance.
Operationally, this converts scheduling into a \textbf{power-allocation} problem: for each workload and SLO, the required power demand is determined, and the scheduler decides \emph{when} and \emph{where} to consume that power.

\begin{figure*}[!t]
	\centering
	\subfigure[Temporal workload]{\includegraphics[width=0.32\linewidth]{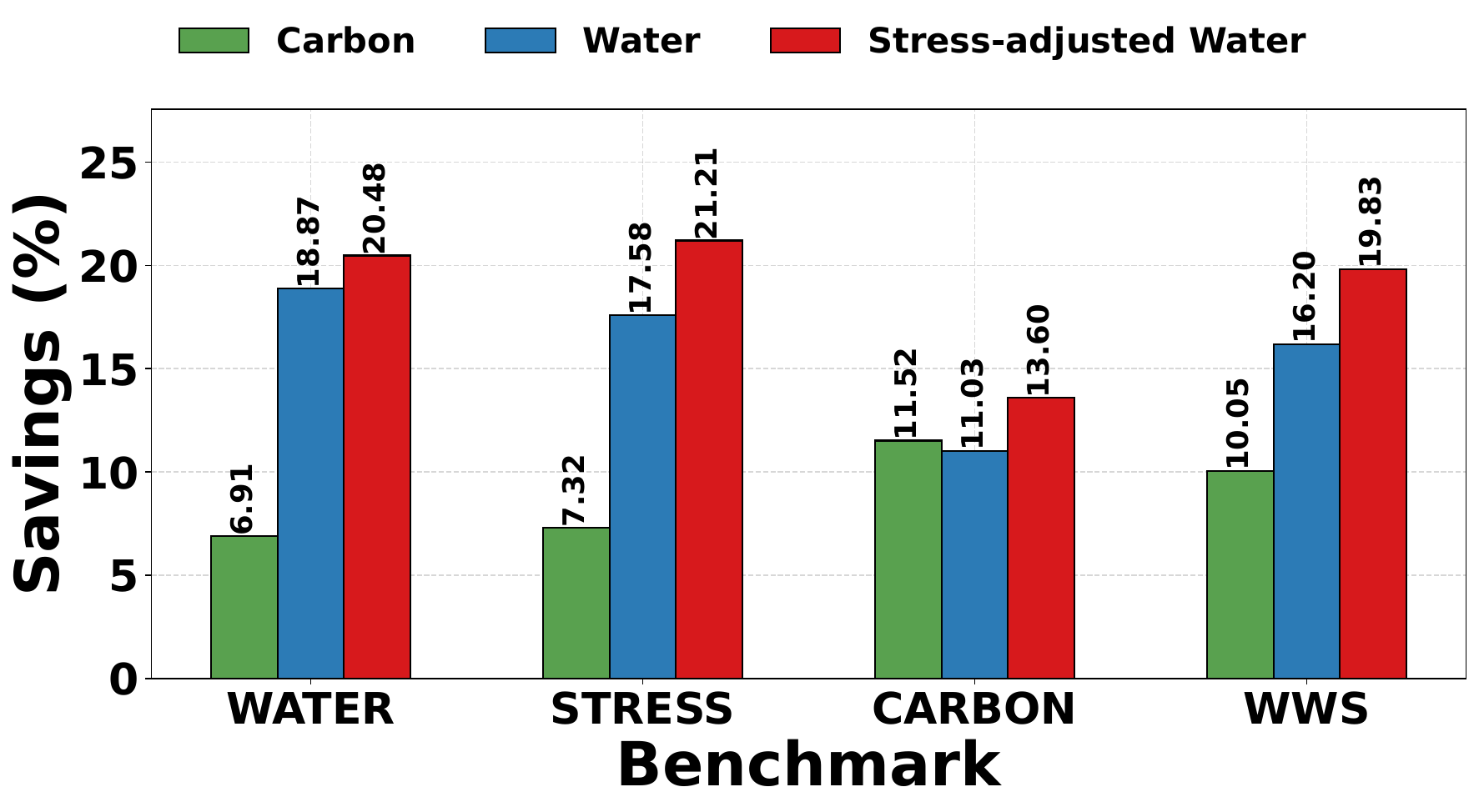}}
	\subfigure[Spatial workload]{\includegraphics[width=0.32\linewidth]{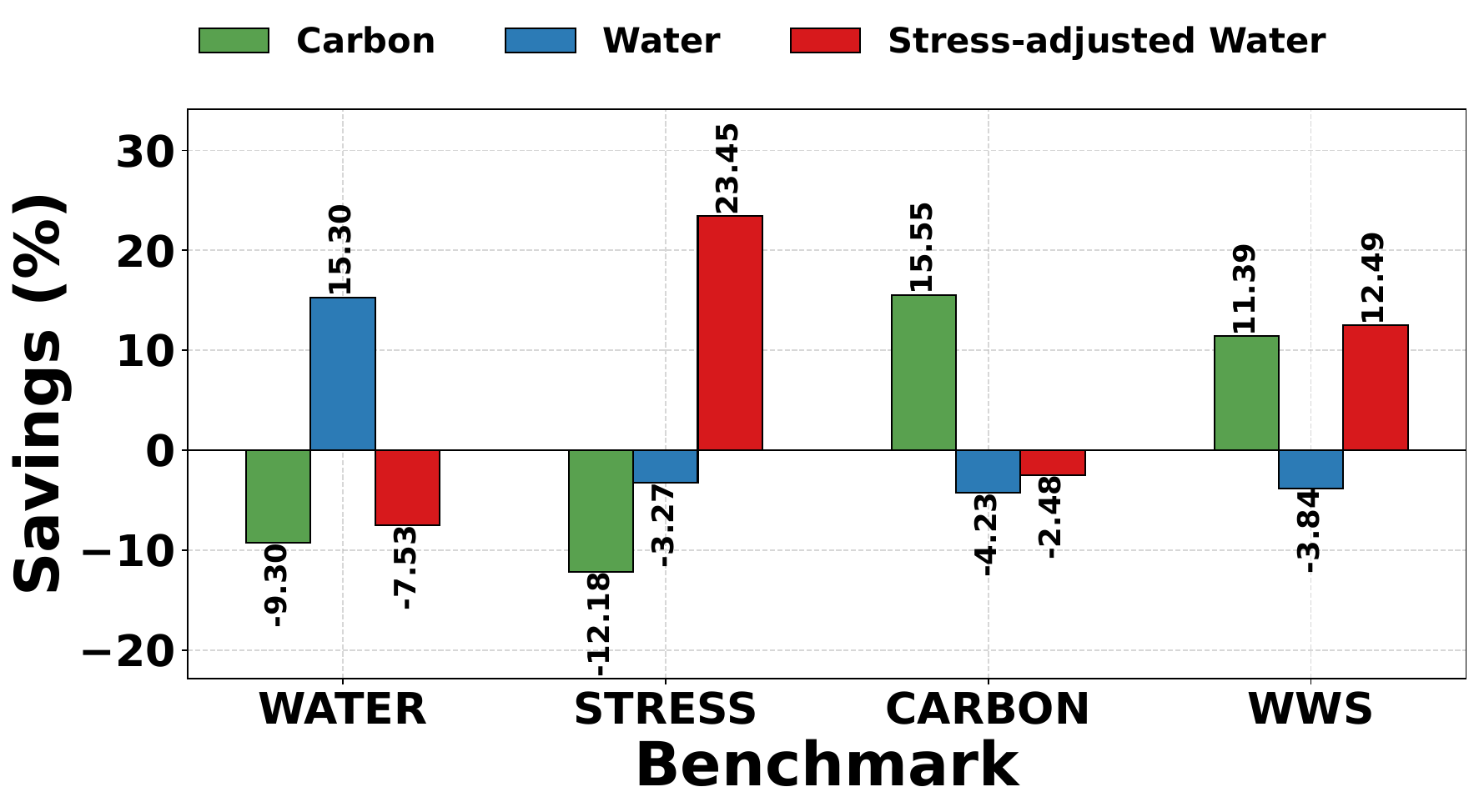}}
	\subfigure[Spatio-temporal workload]{\includegraphics[width=0.32\linewidth]{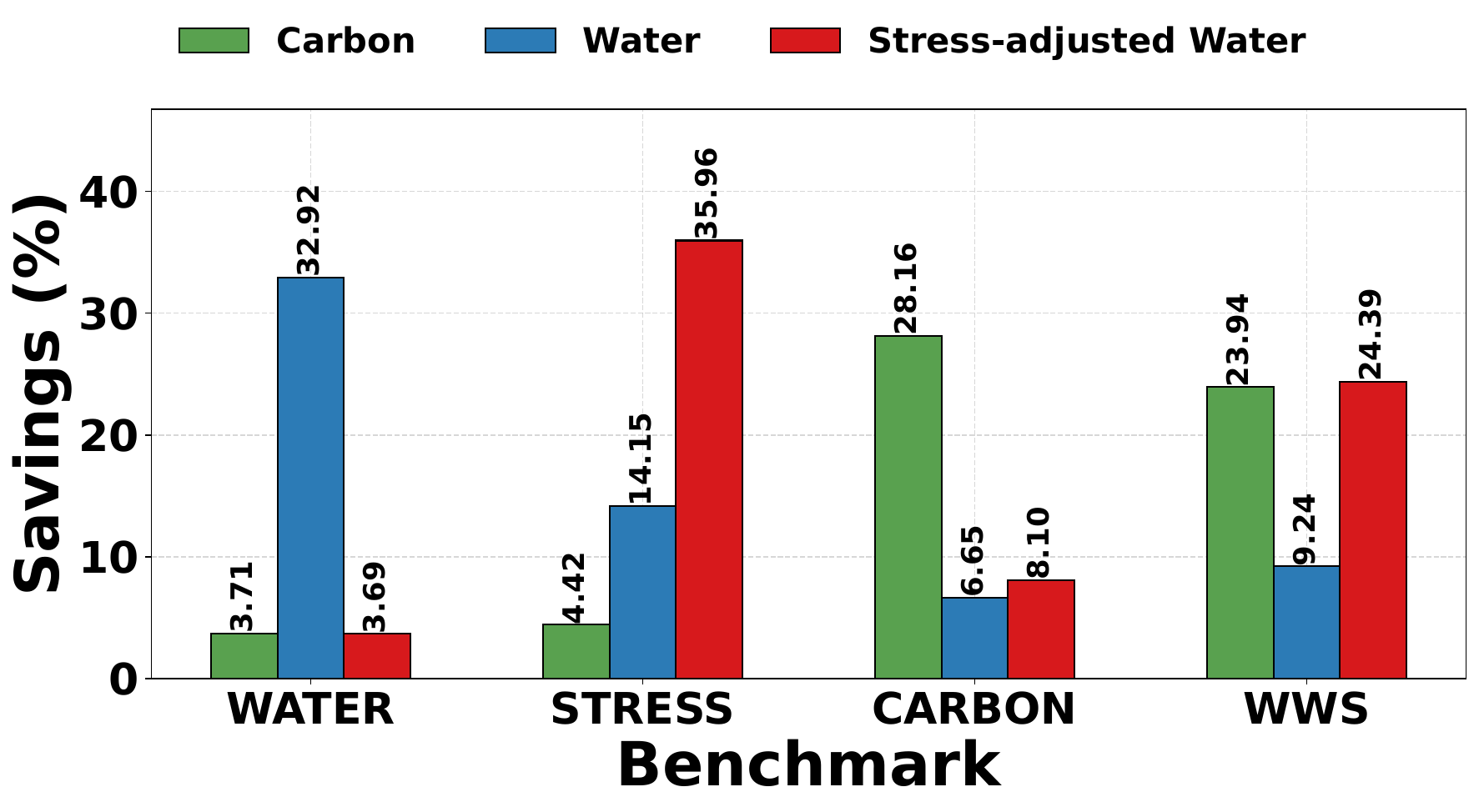}}
	\caption{Savings from workload scheduling under different benchmark strategies, illustrating that minimizing volumetric water can misestimate stress-adjusted water savings—particularly for spatial and spatio-temporal workloads.}
	\label{fig:load_balacing}
\end{figure*}

\subsubsection{Jointly optimizing water and carbon}
Although this paper focuses on water, carbon emissions remain a core sustainability objective.
Importantly, water and carbon are not necessarily aligned.
On-site water depends on weather and cooling dynamics, while off-site water depends on generation water intensity and water stress, which can be high even for low-carbon electricity sources.
Fig.~\ref{fig:water_carbon_variation} illustrates that hourly water and carbon efficiencies can vary independently.
Therefore, optimizing only water or only carbon can unintentionally worsen the other, motivating a joint objective that accounts for both carbon and stress-adjusted water.

\subsubsection{Stress-adjusted water scheduling model}
We model workload scheduling as a time-slotted power-allocation problem over a horizon of $T$ slots.
At each slot, workload arrives at one or more gateways and must be assigned to a data center, either immediately or within a limited deferral window (for delay-tolerant workloads).
Each assignment decision induces IT power consumption at a specific location and time; the resulting total facility power then determines (i) carbon emissions, (ii) on-site cooling water consumption (weighted by local water stress), and (iii) off-site water consumption from electricity generation (weighted by the water stress at power-plant locations).

We jointly minimize total carbon and total stress-adjusted water over the horizon.
A tunable weight parameter controls the emphasis between water and carbon, allowing exploration of the trade-off space.
All workloads must be scheduled, and each data center must respect a maximum power capacity.
Full formal definitions and the complete optimization formulation are provided in Appendix~A.

\textbf{Offline upper bound.}
Optimally scheduling workloads online requires forecasting future carbon intensity, water efficiency, and water stress signals.
Rather than proposing a new online algorithm, our goal is to quantify the \emph{maximum achievable benefit} of workload shifting under idealized conditions.
Accordingly, we compute an offline optimal schedule assuming perfect future knowledge.
This offline solution serves as an upper bound on potential savings and isolates the value of spatiotemporal flexibility independent of forecasting or control errors.

\begin{figure}[!t]
	\centering
	\subfigure[Temporal workload]{\includegraphics[width=0.49\linewidth]{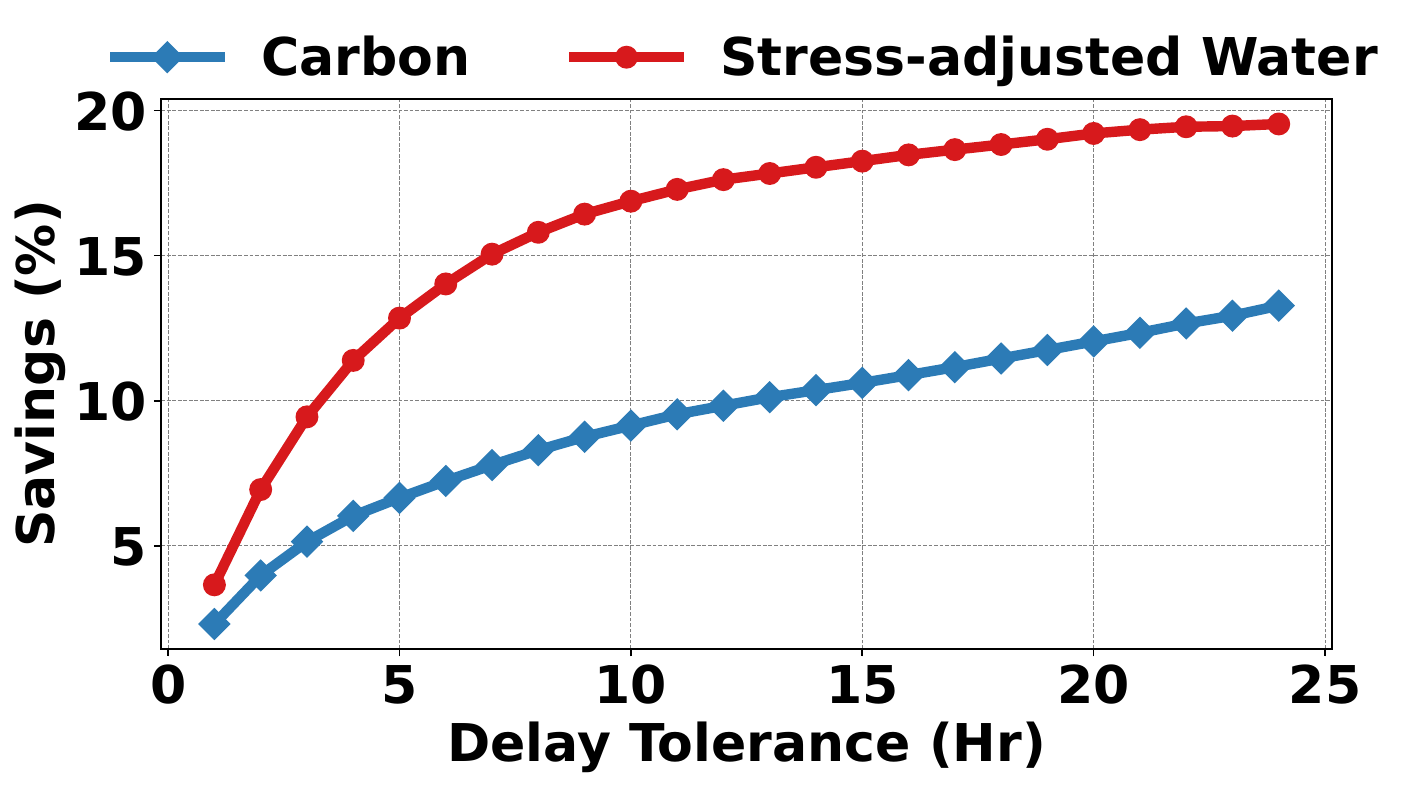}}
	\subfigure[Spatio-temporal workload]{\includegraphics[width=0.49\linewidth]{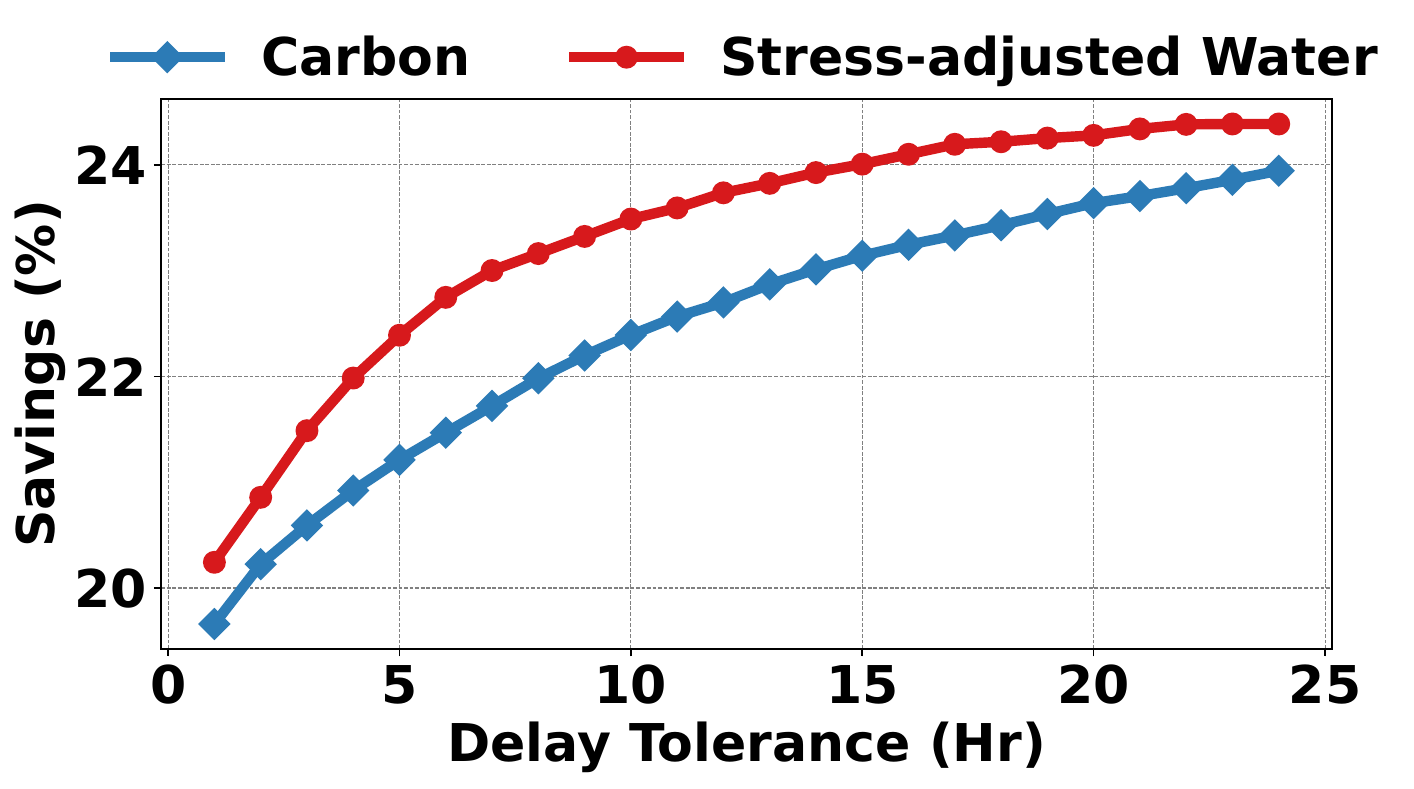}} 
	\caption{Impact of temporal delay tolerance on sustainability savings under temporal and spatio-temporal workload.}
	\label{fig:impact buffer}
\end{figure}

\subsubsection{Evaluation setup}
We evaluate stress-adjusted water scheduling using long-term trace-driven simulations.
We consider the top five U.S. data center markets by concentration and extend the analysis to the top twenty.
We use hourly on-site/off-site water efficiencies and carbon efficiencies from \cite{gupta2024dataset}.
We use county-level monthly water stress factors from AWARE-US \cite{lee2019aware}, and off-site stress factors derived from our eGRID-based methodology (Fig.~\ref{fig:egrid_poc_figure}).
For workload, we use Google search workload traces from 2023 \cite{GoogleTransparencyTraffic} and scale them to represent a 10~MW data center.
We emulate temporal, spatial, and spatio-temporal flexibility by varying the deferral window and the set of reachable markets.
Geographic distance between markets is used to model routing overhead for interactive workloads.

\textbf{Baselines.}
We compare our joint optimization \ouralg (Water Wise Scheduling) that minimizes stress-adjusted water plus carbon against:
(i) \textbf{NoScheduling} that processes workload immediately in the local data center,
(ii) \carbon that minimizes only the carbon emission,
(iii) \water that minimizes only the water \emph{volume},
and (iv) \eWater that minimizes only the stress-adjusted water.

\subsubsection{Results}
\textbf{Overall savings and misestimation under different workload flexibilities.}
Fig.~\ref{fig:load_balacing} summarizes savings under temporal, spatial, and spatio-temporal workloads across different optimization objectives.
Temporal flexibility primarily exploits intra-market variation, and consequently volumetric water optimization provides a reasonable—but still incomplete—approximation of stress-adjusted outcomes.

In contrast, for \emph{spatial} and \emph{spatio-temporal} workloads, optimizing for water volume alone can substantially misrepresent true water impact.
As shown in Fig.~\ref{fig:load_balacing}(b) and (c), strategies that minimize volumetric water often achieve modest water savings while simultaneously increasing stress-adjusted water.
This occurs because spatial routing shifts workloads toward regions with lower cooling water usage but significantly higher water stress or stressed electricity supply, amplifying off-site stress-adjusted water consumption.

By jointly accounting for on-site and off-site water stress, \ouralg avoids these failure modes.
Across all workload types, it consistently achieves large reductions in stress-adjusted water while also reducing carbon emissions, avoiding the ``optimize-one-harm-the-other'' behavior exhibited by single-objective baselines.
These results demonstrate that volumetric water savings alone are insufficient to assess sustainability, particularly when workloads are geographically flexible.

\begin{figure}[t!]
	\centering
	\subfigure[Spatial workload]{\includegraphics[width=0.49\linewidth]{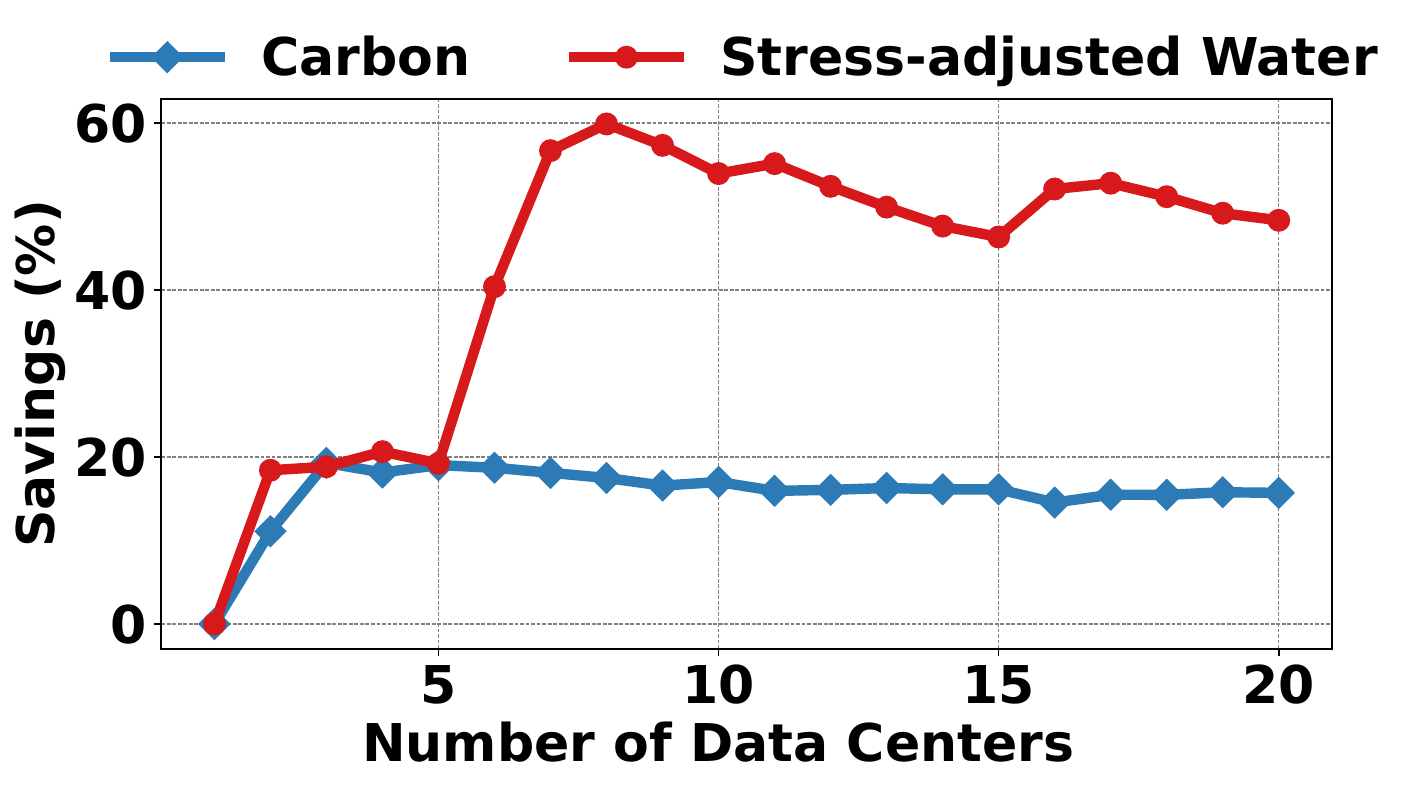}}
	\subfigure[Spatio-temporal workload]{\includegraphics[width=0.49\linewidth]{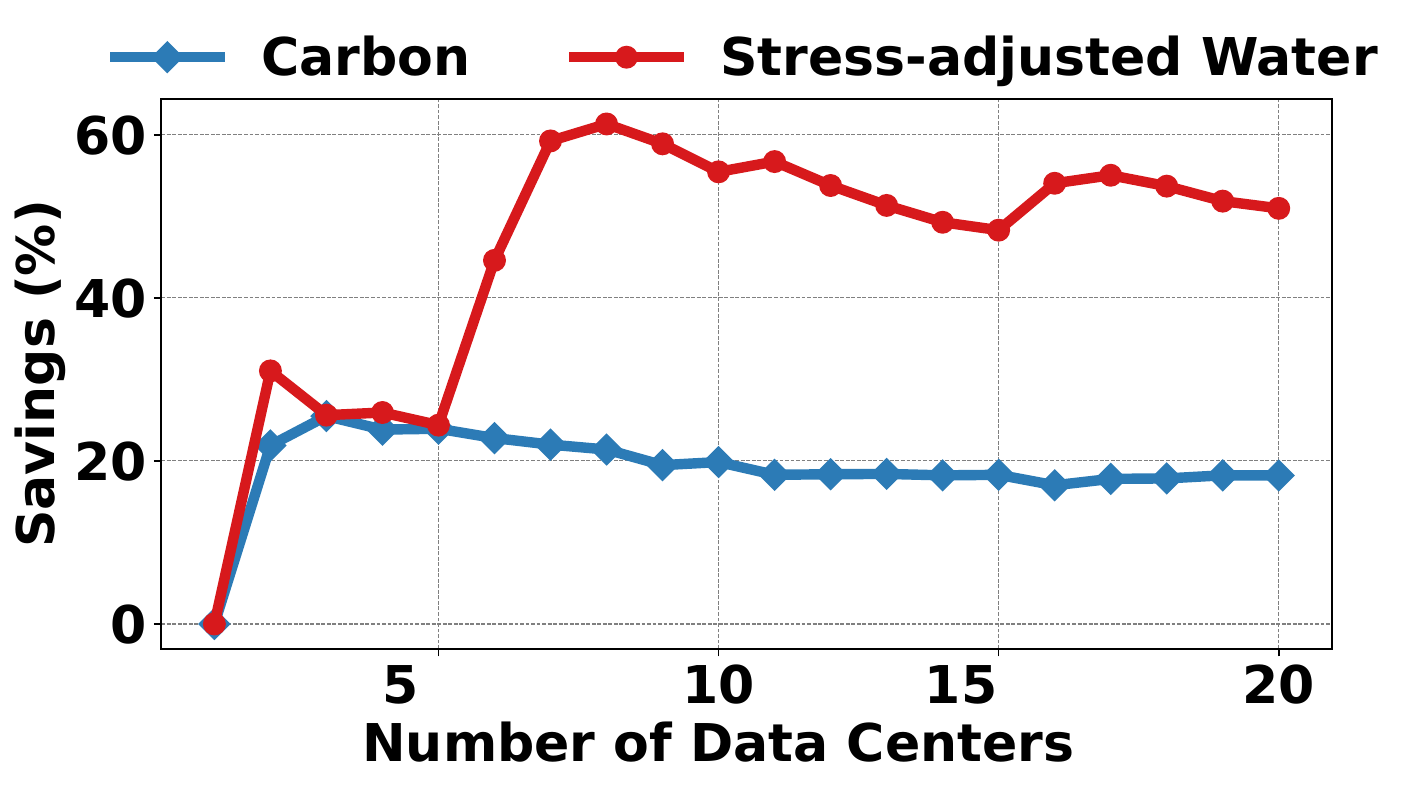}}
	\caption{Impact of the number of available data centers.}
	\label{fig:impact data center num}
\end{figure}

\textbf{Impact of delay tolerance.}
We vary the delay tolerance from 1 to 24 hours (Fig.~\ref{fig:impact buffer}).
For temporally flexible workloads, longer deadlines yield higher savings by expanding the set of feasible low-impact hours.
For spatio-temporal workloads, even a small delay tolerance provides strong benefits because the scheduler can combine modest temporal shifts with geographic diversity.
Overall, spatio-temporal flexibility provides substantial gains even at low delay tolerance, while purely temporal flexibility benefits most from long deadlines.

\textbf{Impact of the number of data centers.}
We vary the number of available markets from 1 to 20.
Savings increase rapidly as additional markets introduce greater spatiotemporal heterogeneity, then exhibit diminishing returns as marginal diversity decreases.

\begin{figure}[t!]
	\centering
	\subfigure[Temporal workload]{\includegraphics[width=0.49\linewidth]{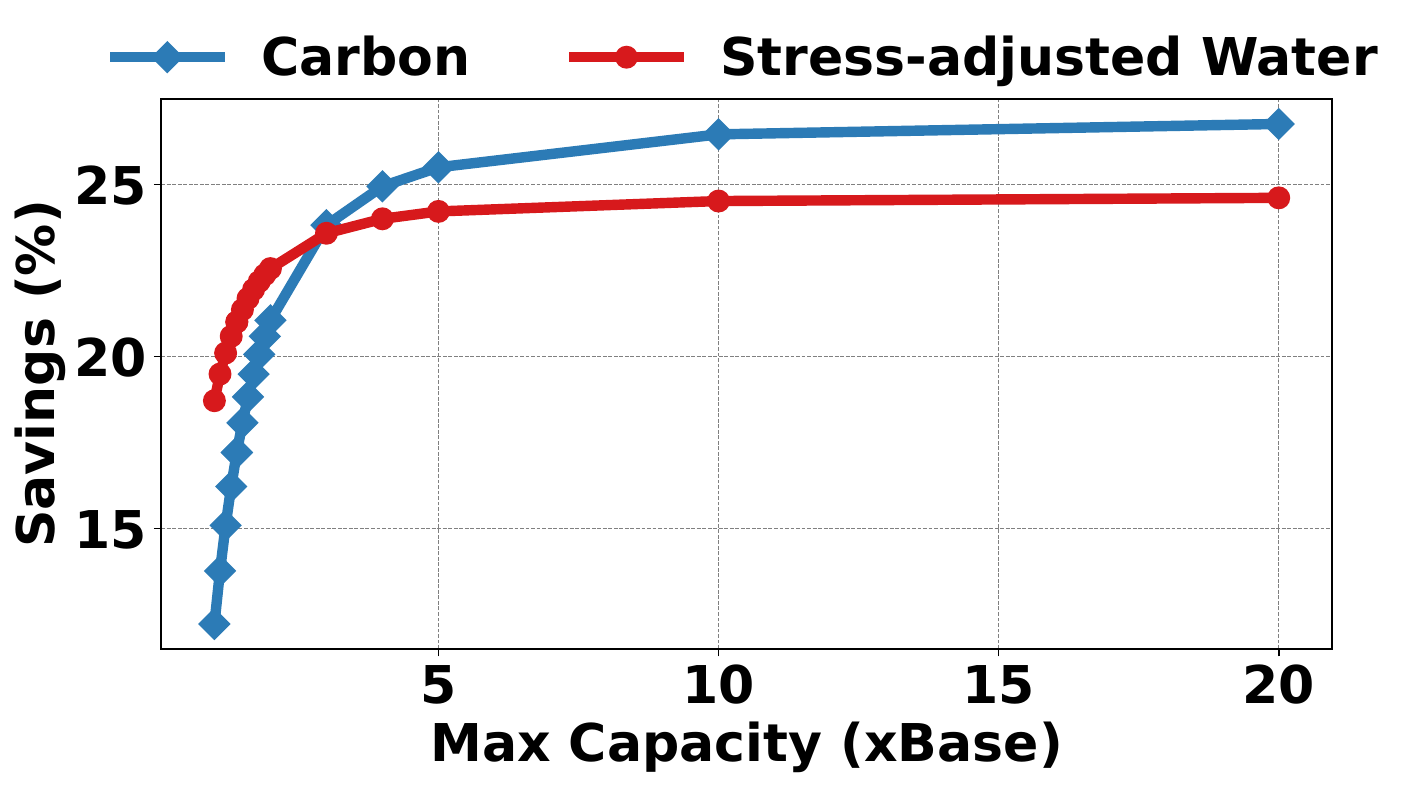}}
	\subfigure[Spatio-temporal workload]{\includegraphics[width=0.49\linewidth]{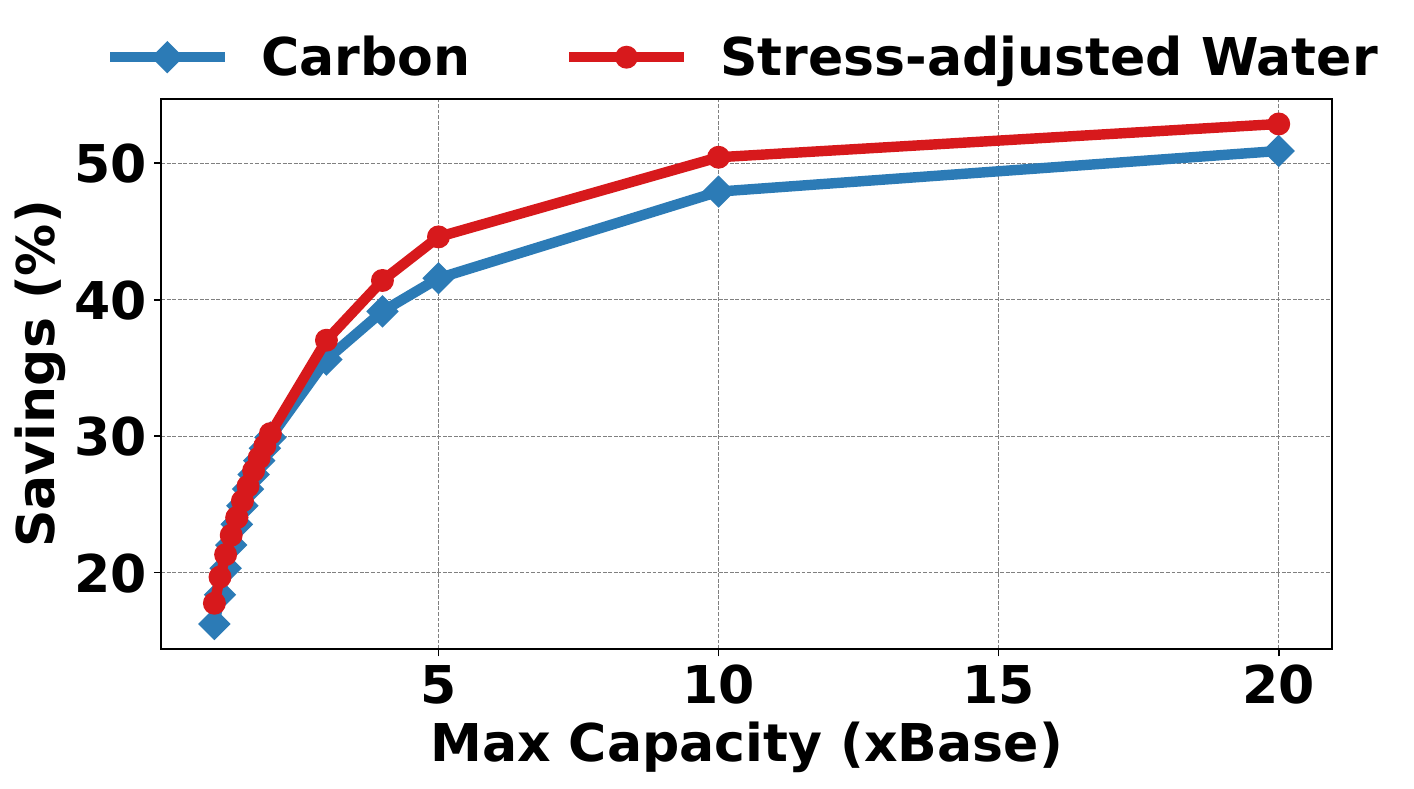}}
	\caption{Impact of data center capacity headroom.}
	\label{fig:impact data center capacity}
\end{figure}

\textbf{Impact of capacity headroom.}
We scale data center capacity up to $20\times$ the baseline.
Additional capacity enables workloads to be concentrated into lower-impact times and locations, improving both carbon and stress-adjusted water outcomes.
However, benefits saturate beyond moderate scaling, indicating that heterogeneity and flexibility—rather than unlimited capacity—are the primary drivers of sustainability gains.

\subsection{Rainwater Harvesting}
\label{sec:rain}

\begin{figure}[t!]
	\centering
	\includegraphics[width=1\linewidth]{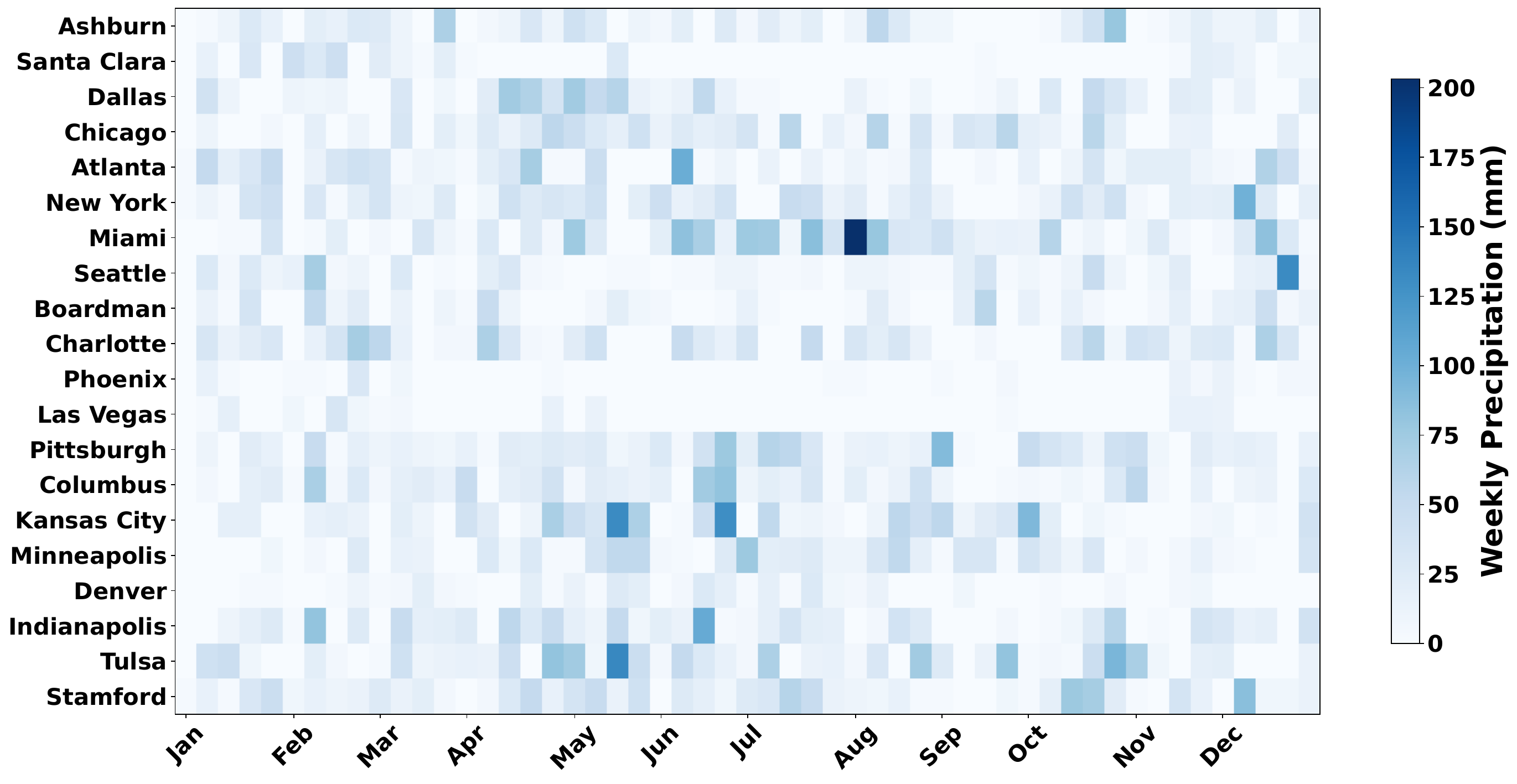}
	\caption{Annual precipitation patterns across major U.S. data center markets, illustrating strong spatial and seasonal variability.}
	\label{fig:precip_heatmap}
\end{figure}

Rainwater harvesting provides a complementary lever to offset on-site withdrawals by substituting municipal supply with locally captured precipitation.
Beyond offsetting freshwater demand, rainwater harvesting can mitigate stormwater runoff pollution and alleviate pressure on sewage infrastructure \cite{Campisano2012}.
Several U.S. states encourage rainwater harvesting through incentives and rebates, particularly in drought-prone regions \cite{TexasWater2020}.
In data centers, harvesting can also support sustainability certifications, reduce exposure to water price volatility, and improve resilience under drought restrictions \cite{USGBC2019}.
Rainwater typically requires minimal treatment (e.g., filtration) for cooling use; moreover, its lower mineral content can improve cycles of concentration, increasing the fraction that can be consumed via evaporative cooling \cite{Ghisi2007,rahmani2017reducing}.
Major operators have deployed harvesting systems in practice \cite{Google2022,Equinix2021}. Fig.~\ref{fig:precip_heatmap} shows the variability in precipitation across the major U.S. data center markets throughout the year.

\textbf{System model.}
A harvesting deployment consists of (i) a collection surface (roofs, parking lots, and potentially nearby buildings) and (ii) storage (tanks or retention ponds).
Effectiveness depends on rainfall seasonality, storage sizing, and available collection area.
Accordingly, we quantify feasibility across major U.S. markets as a function of harvesting area and storage capacity.

\begin{figure*}[t!]
	\centering
	\subfigure[0.5 million gallons]{\includegraphics[width=0.32\linewidth]{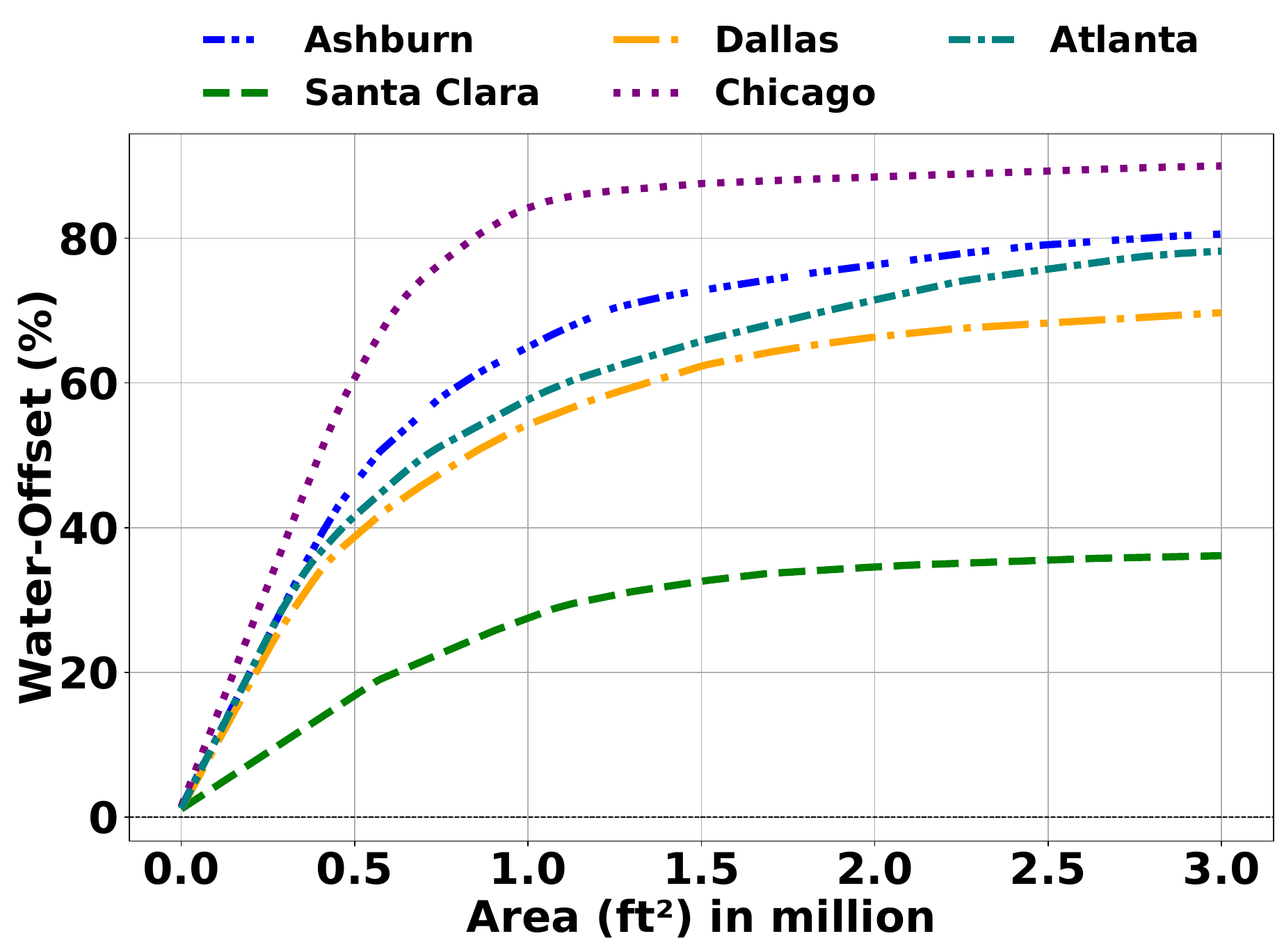}}
	\subfigure[1 million gallons]{\includegraphics[width=0.32\linewidth]{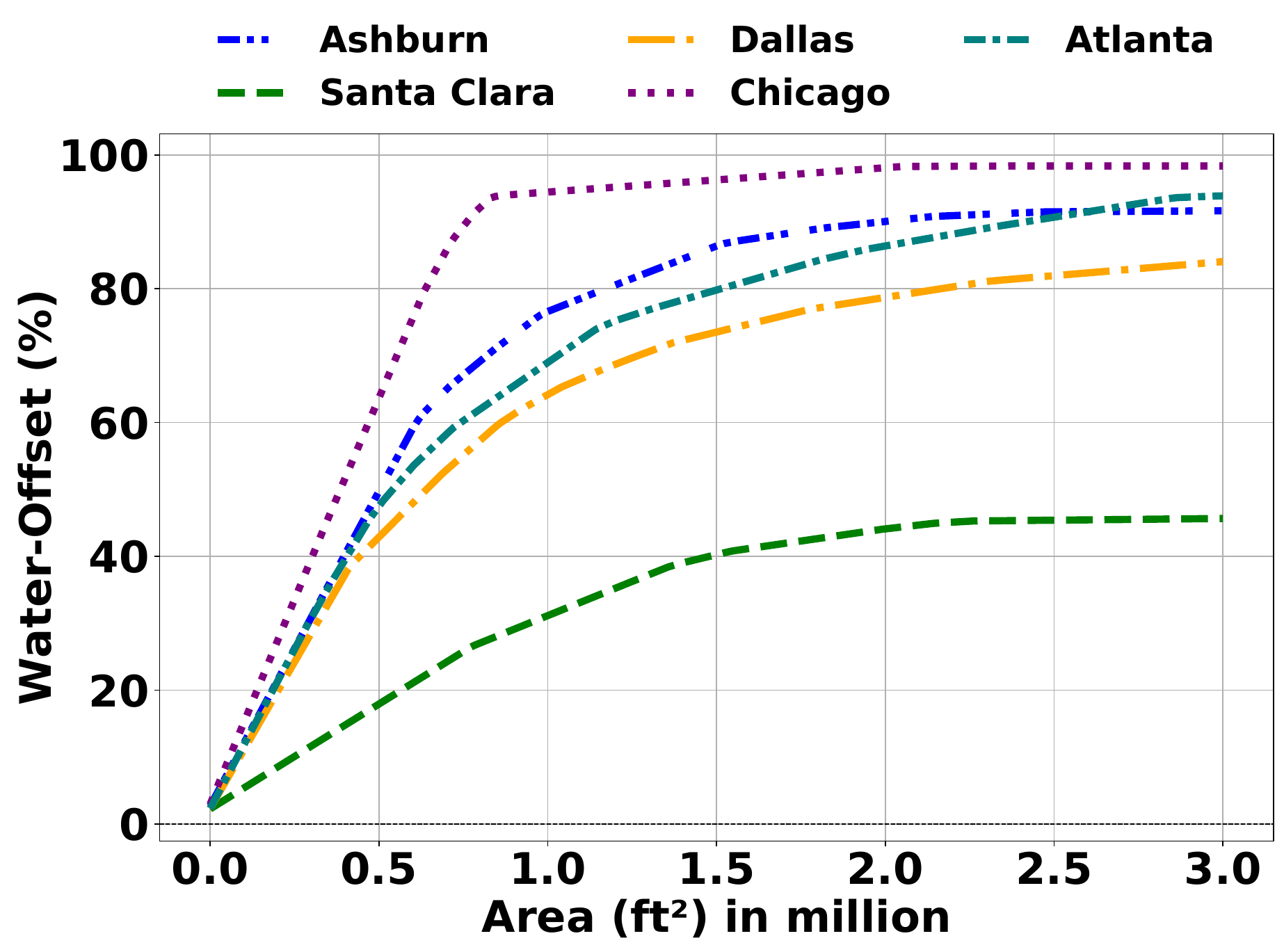}}
	\subfigure[3 million gallons]{\includegraphics[width=0.32\linewidth]{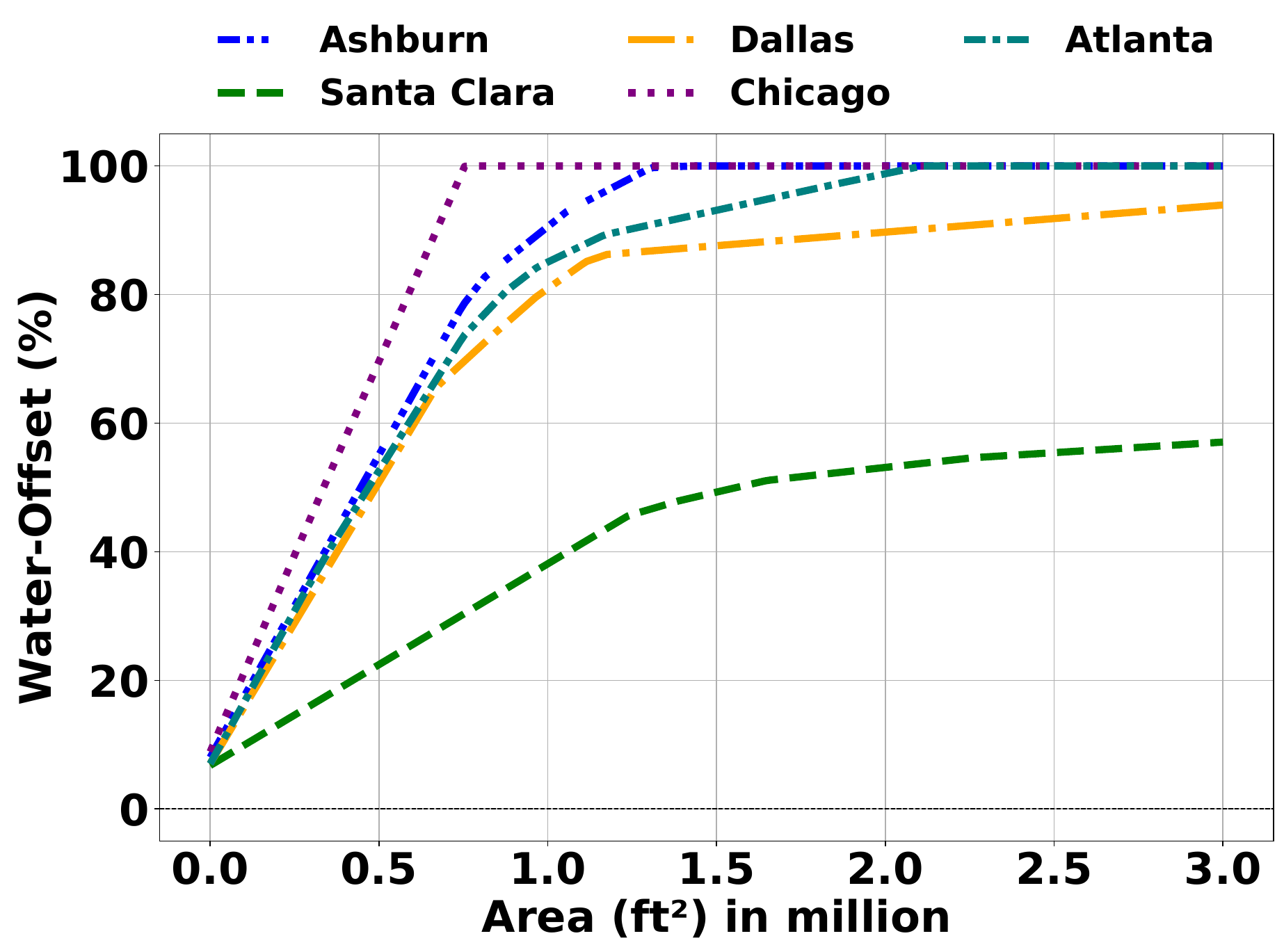}}
	\caption{Water-offset (water saving due to rainwater use) versus harvesting area under different tank sizes across major data center markets.}
	\label{fig:sustainability vs rain harvest}
\end{figure*}

\begin{figure}[t!]
	\centering
	\includegraphics[width=1\linewidth]{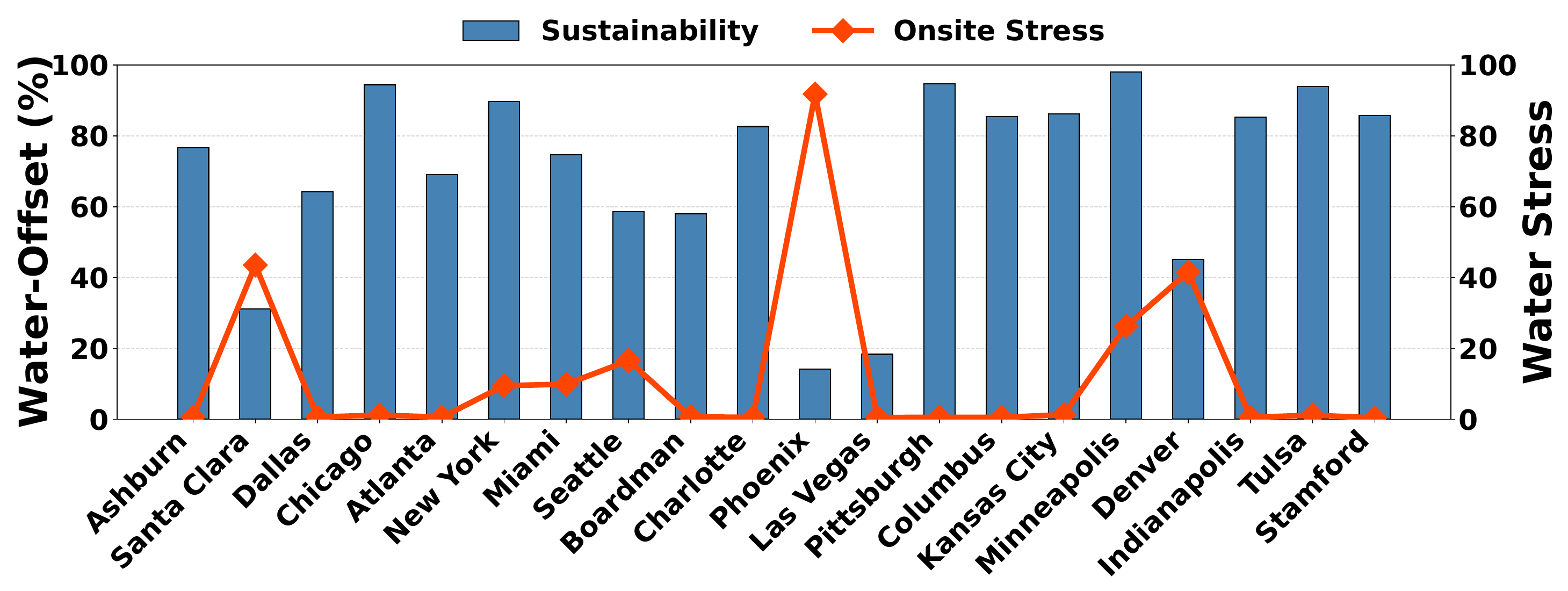}
	\caption{Rainwater offset for a 1{,}000{,}000-gallon tank and 1{,}000{,}000~sqft harvesting area (top markets), with corresponding stress context.}
	\label{fig:sustainability all city}
\end{figure}

\subsubsection{Evaluation setup}
We run a year-long simulation for major data center markets using hourly precipitation data from Weather Underground \cite{weatherUnderground}.
We use the same 10~MW workload scaling as in scheduling, with hourly IT load ranging from 30\% to 100\% of peak.
We assume tanks start half full and adopt a cycle-of-concentration of 10, meaning approximately 90\% of the harvested water can be used for evaporative cooling.

We consider harvesting areas from 300{,}000 to 1{,}500{,}000~sqft, representing (i) roof-only collection and (ii) expanded surfaces including parking and adjacent buildings.
We consider tank sizes from 500{,}000 to 3{,}000{,}000 gallons.
We report \textbf{sustainability}, defined as the fraction of total annual cooling demand supplied by harvested rainwater.

\subsubsection{Results}
Fig.~\ref{fig:sustainability vs rain harvest} shows water-offset versus harvesting area for fixed tank sizes.
Larger harvesting surfaces generally increase water offset, but benefits saturate once rainfall supply exceeds storage or demand.
Increasing tank size improves buffering and reduces overflow loss, particularly in markets with strong seasonality.

Fig.~\ref{fig:sustainability all city} summarizes water offset across the top markets for a representative configuration (1{,}000{,}000-gallon tank and 1{,}000{,}000~sqft harvesting area).
Most markets have high potential for water offset through rainwater harvesting, while arid markets such as Phoenix and Denver have lower potential due to limited precipitation.
Importantly, these low water-offset markets often coincide with high water stress, increasing the value of each liter saved and reinforcing the need for stress-aware evaluation.

\subsection{Dry Cooling}
\label{sec:dry}

\begin{figure}
	\centering
	\includegraphics[trim = 6cm 8cm 0cm 3cm, clip, width=1\linewidth]{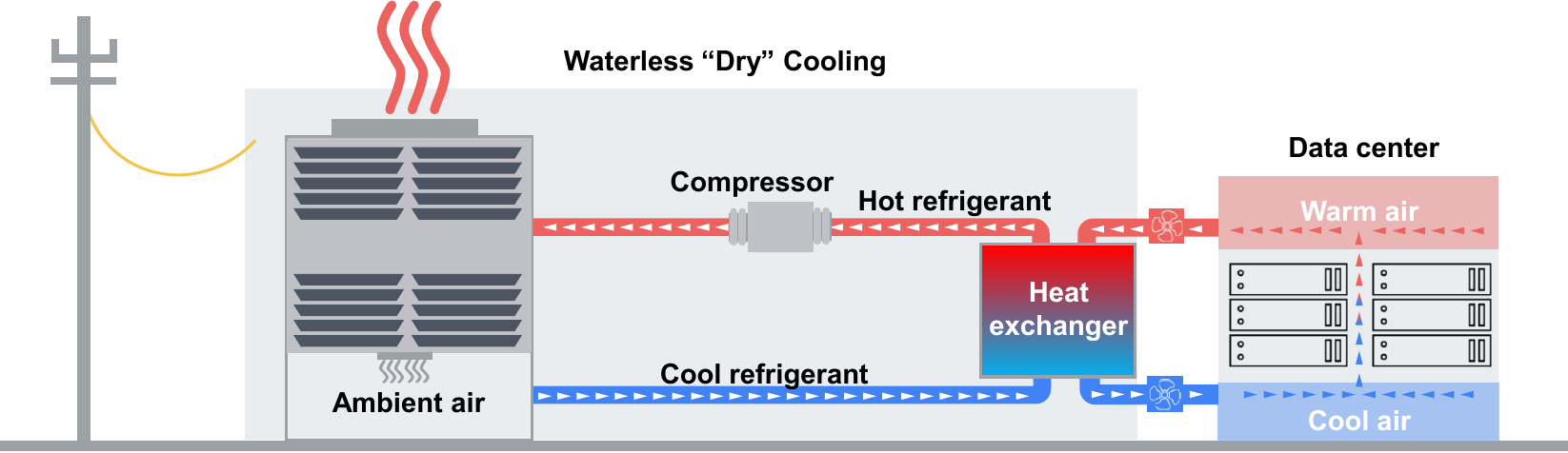}
	\caption{Illustration of dry cooling in data centers.}
	\label{fig:air_cooling}
\end{figure}

Dry cooling eliminates \textbf{on-site} cooling water by rejecting heat through air-based systems (e.g., air-cooled chillers or dry coolers), as illustrated in Fig.~\ref{fig:air_cooling}. Advanced high-efficiency cooling systems often use a hybrid approach, combining chillers with free-air cooling by directly (after particulate filtering) pushing outside air into server rooms when it is cold enough (e.g., $<27^o$C).
Dry cooling is attractive in water-scarce regions, but it typically increases energy consumption and, in turn, \textbf{off-site} water consumption and carbon emissions from electricity generation.
Therefore, dry cooling is best understood as shifting water impact from Scope~1 to Scope~2 rather than eliminating it.

\subsubsection{Evaluation setup}
We evaluate dry cooling across major U.S. markets using year-long simulations.
We use the same workload traces and hourly location-specific on-site/off-site water and carbon efficiencies as before.
We vary PUE from 1.2 (baseline, efficient water-based evaporative cooling) to 2.0 to represent an increasing energy penalty with dry cooling.
Under dry cooling, on-site water consumption is set to zero, and off-site impacts scale with increased electricity use.
This sweep captures a wide range of possible efficiency penalties observed across climates and system designs.

\subsubsection{Results and implications}
\begin{figure}[!t]
	\centering
	\subfigure[Water savings]{\includegraphics[width=0.48\linewidth]{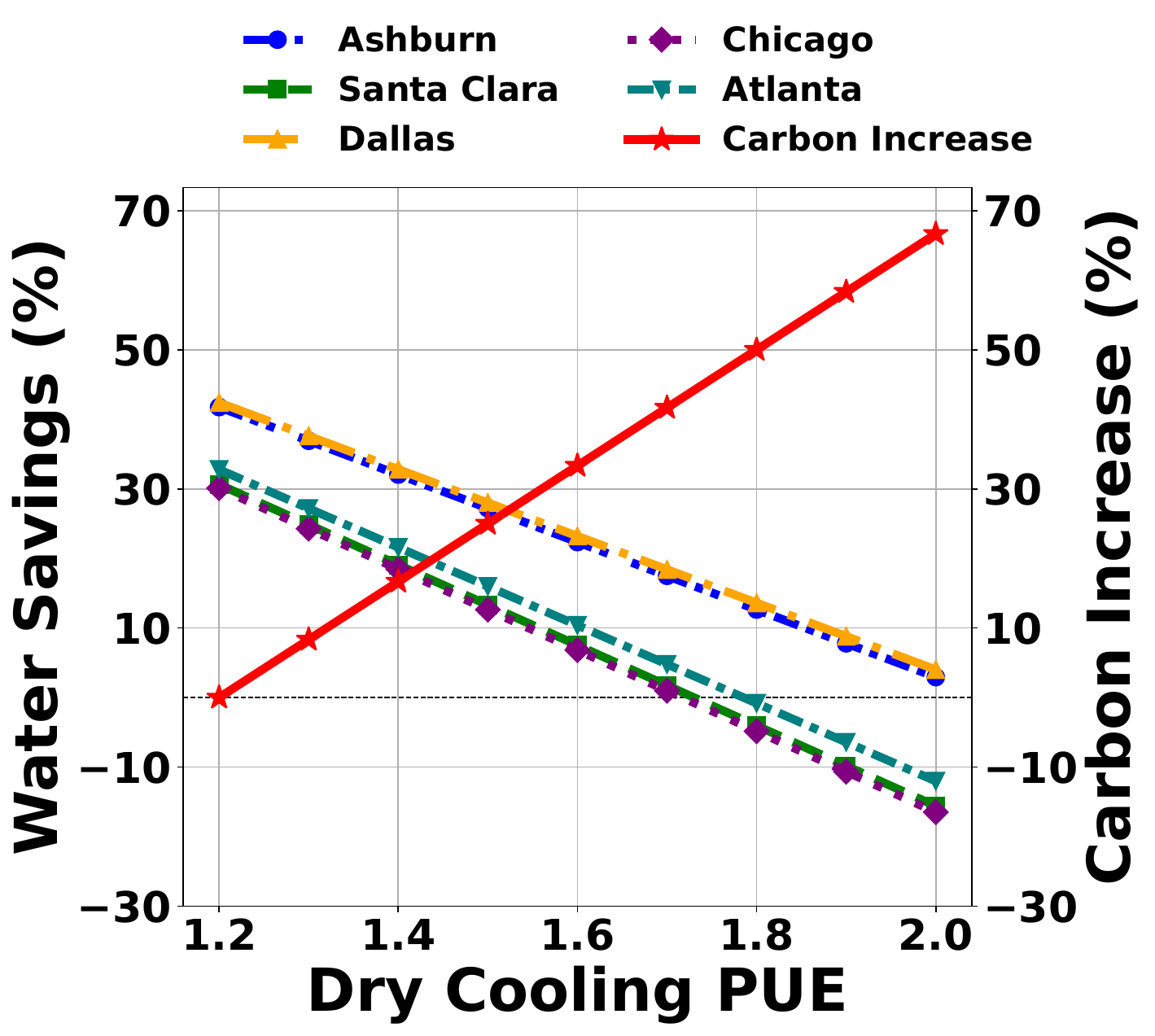}
		\label{fig:water_savings_normal}}
	\subfigure[Stress-adjusted water savings]{\includegraphics[width=0.48\linewidth]{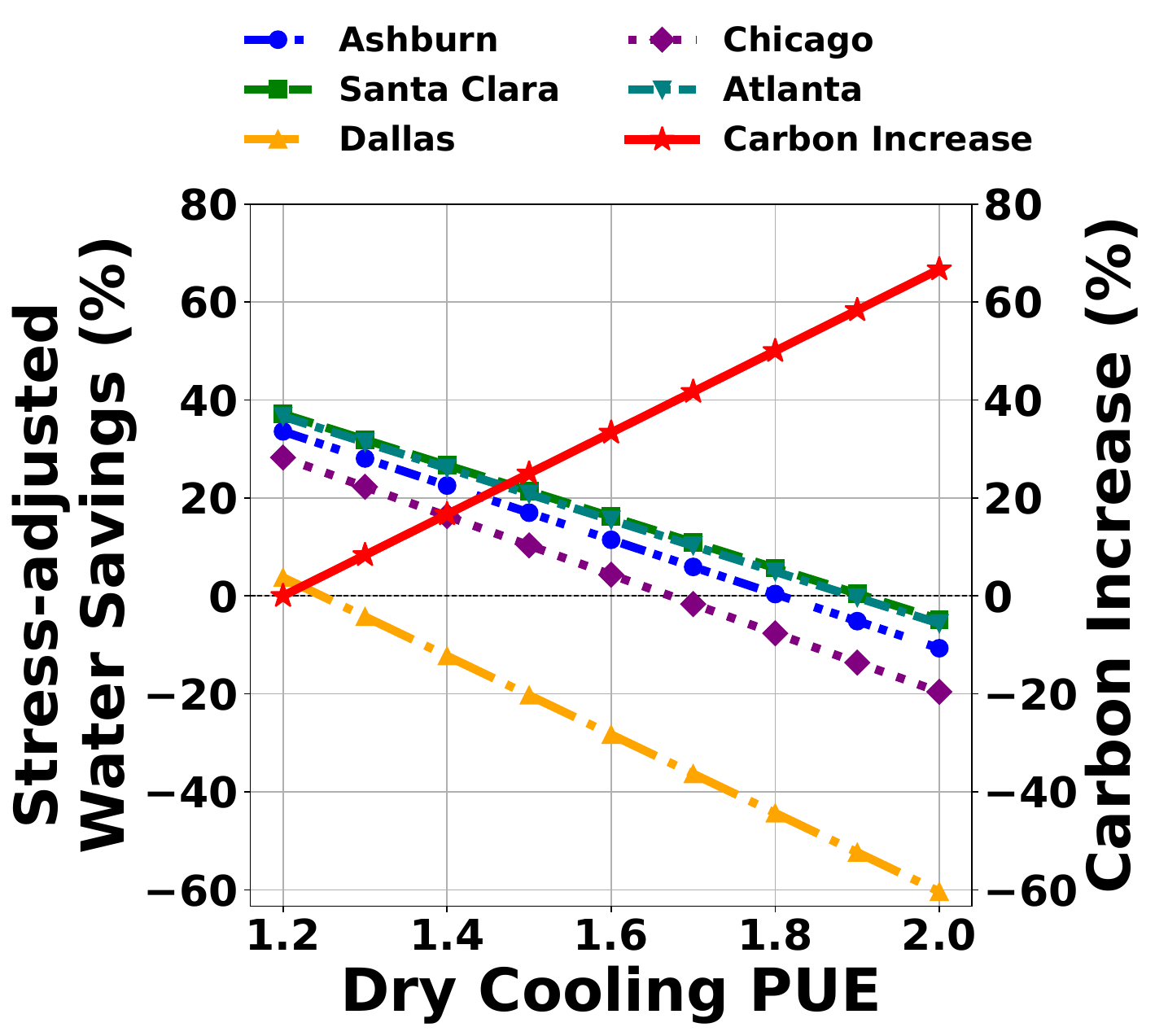}
		\label{fig:water_savings_stress}}
	\caption{Trade-offs between water and carbon impacts under dry cooling as PUE increases (baseline water cooling PUE = 1.2).}
	\label{fig:water_savings}
\end{figure}

Fig.~\ref{fig:water_savings} shows water savings and stress-adjusted water savings as PUE increases.
We also look at the increase in carbon (from a baseline PUE of 1.2) due to higher grid-power consumption resulting from the higher PUE. Carbon emissions increase approximately linearly with PUE, reflecting proportional increases in electricity consumption.
At low PUE (e.g., 1.3--1.5), dry cooling can yield substantial reductions in total water volume by eliminating on-site evaporation.
However, as PUE increases, off-site water and carbon penalties grow, potentially offsetting or negating the water benefit.

Stress-adjusted savings reveal a critical nuance: markets can differ substantially in whether dry cooling reduces or increases \emph{stress-adjusted} water because the off-site stress depends on the supplying eGRID region.
Consequently, evaluating dry cooling using water volume alone can be misleading.

\begin{figure*}
	\centering
	\includegraphics[width=0.7\linewidth]{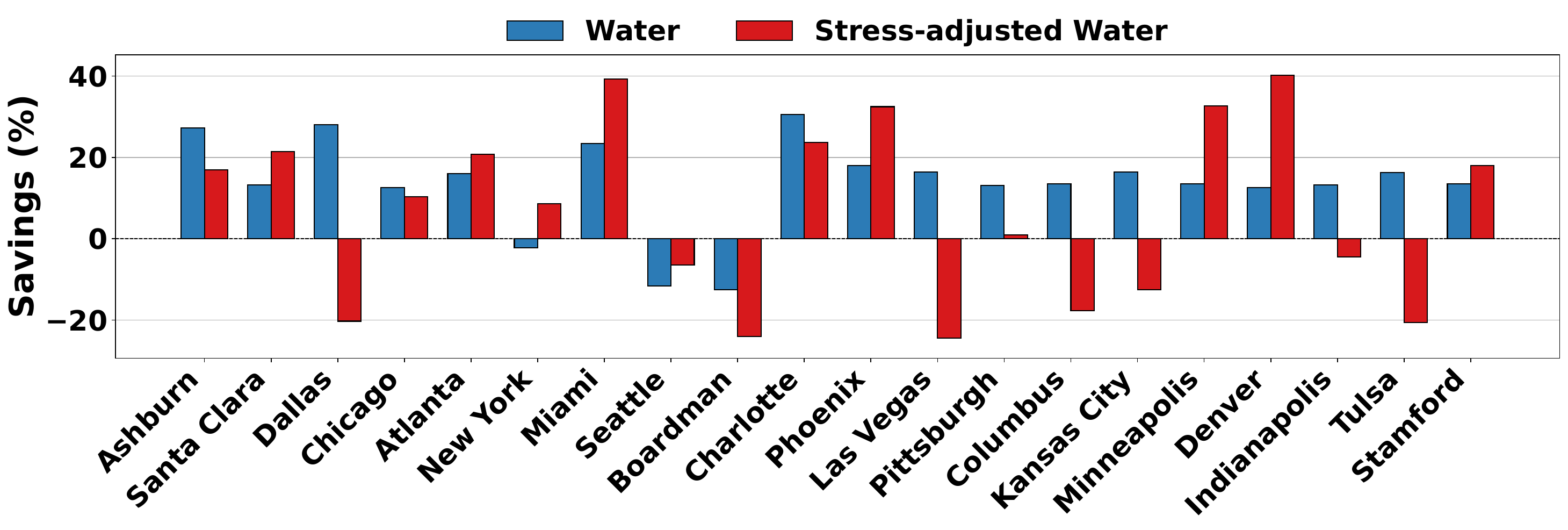}
	\caption{Dry cooling impact across major markets at PUE=1.5: volumetric water savings versus stress-adjusted water change.}
	\label{fig:water_savings all city}
\end{figure*}

To understand broader market impact, Fig.~\ref{fig:water_savings all city} fixes PUE at 1.5 and compares water and stress-adjusted water changes across major markets.
While many markets show positive volumetric water savings, stress-adjusted results can flip sign in regions where electricity is supplied by highly stressed generation, as increased electricity demand amplifies off-site impacts.

\textbf{Feasibility guidance.}
Dry cooling is most attractive when its efficiency penalty is small (e.g., PUE $\le$ 1.5), in which case eliminating on-site evaporation outweighs off-site penalties.
At higher PUE regimes (e.g., $\ge$ 1.8), savings diminish, and carbon penalties become large, making dry cooling less attractive unless paired with grid decarbonization or other mitigating measures.
Overall, dry cooling should be deployed with a stress-aware lens: it can be highly beneficial in some high-stress markets but counterproductive in markets where electricity is sourced from water-stressed regions.

\textbf{When is dry cooling beneficial?}
Dry cooling is recommended when the reduction in on-site stress-adjusted water exceeds the increase in off-site stress-adjusted water induced by higher electricity demand.
In practice, this occurs when local water stress is high, and the supplying grid’s water stress is relatively low.

\section{Related Work}

\textbf{Sustainability-aware data center operation.}
Prior work has extensively studied sustainability-aware data center operation by adapting workload placement, scaling, and resource allocation to environmental signals.
Most systems focus on carbon-aware optimization, exploiting temporal variation in grid carbon intensity \cite{hanafy2023carbonscaler,wiesner2021let} and geographic diversity across data centers \cite{liu2011greening,maji2023bringing}, or combining both dimensions \cite{sukprasert2024limitations}.
These approaches are effective for carbon, whose impacts are global, but implicitly assume spatially homogeneous environmental effects—an assumption that does not hold for water.

\textbf{Water metrics and accounting.}
WUE is the standard metric for reporting data center water efficiency \cite{azevedo2011water}, but it captures water volume rather than environmental impact.
Recent efforts recognize this limitation: SCARF introduces an AWI metric that weights water use by regional scarcity \cite{wu2025not}, demonstrating that water-aware accounting can substantially alter sustainability conclusions.
However, SCARF applies regional adjustments at an aggregate level and does not distinguish between on-site and off-site water pathways or model electricity-generation provenance.

\textbf{Water-aware scheduling and trade-offs.}
Early systems reduce total water consumption by shifting spatial and temporal workloads \cite{Islam2014WATCH,islam2015water,islam2016exploiting}, but optimize volumetric water use without accounting for stress.
More recently, WaterWise incorporates water stress into scheduling decisions and shows that optimizing water and carbon independently can lead to conflicting outcomes \cite{jiang2025waterwise}.
However, it models off-site water impacts using region-level abstractions and does not account for the geographic locations of water-consuming power plants, leading to significant misestimation of off-site water stress.

\textbf{Infrastructure-based techniques.}
Complementary work studies cooling-system design and water-saving infrastructure, including flexible cooling \cite{gnibga2024flexcooldc} and rainwater harvesting \cite{ahmed2014can}.
These studies highlight strong location dependence but typically evaluate volumetric water savings.
Our work differs by evaluating both software- and infrastructure-level techniques using \emph{stress-adjusted water}, explicitly accounting for spatiotemporal water value and on-site/off-site impacts.

% \textbf{Summary.}
% In summary, prior work has made significant progress in reducing the energy, carbon, and water footprints of data centers. However, most existing approaches treat water as a homogeneous resource or rely on coarse regional abstractions for offsite impacts. Our work fills this gap by introducing stress-adjusted water accounting that jointly captures onsite seasonality and offsite provenance, enabling more accurate evaluation of scheduling, rainwater harvesting, and dry-cooling strategies.

\section{Conclusion}

As data centers continue to grow in scale and importance, their environmental footprint—particularly water consumption—has become an increasingly critical yet under-addressed sustainability challenge. Unlike carbon emissions, whose impacts are largely global, water impacts are inherently local, seasonal, and spatially heterogeneous. This paper advances the notion of \emph{stress-adjusted water}, arguing that sustainable data center operation must account not only for how much water is consumed, but also for \emph{where} and \emph{when} that consumption occurs, and whether it arises on-site or through electricity generation.

We introduce a stress-adjusted water accounting framework based on the AWARE-US model that jointly captures on-site and off-site water stress and explicitly models the provenance of electricity consumption. Our analysis reveals substantial regional variation in electricity-water intensity, showing that commonly used fixed national factors can significantly misestimate off-site and stress-adjusted water impacts. Using this framework, we demonstrate that software-based workload scheduling can reduce stress-adjusted water while simultaneously lowering carbon emissions, particularly when spatiotemporal flexibility is available.

Beyond scheduling, we evaluate complementary infrastructure-level interventions. We show that rainwater harvesting can provide an effective supplemental water source in precipitation-rich regions, achieving near-complete or even full on-site water offset under realistic storage constraints. In contrast, dry cooling shifts water impact from on-site to off-site and can increase stress-adjusted water and carbon when the associated energy penalty is high or electricity is sourced from water-stressed regions.

Overall, our findings demonstrate that meaningful improvements in data center water sustainability are achievable through stress-aware scheduling, alternative water sourcing, and carefully evaluated cooling technologies. More broadly, this work underscores the need to treat water as a first-class sustainability metric and to move beyond uniform accounting toward context-aware, impact-driven evaluation of digital infrastructure.

\begin{acks}
	This work is supported in part by the U.S. National Science Foundation under grant numbers ECCS-2152357, CCF-2324915, CCF-2324916, and CCF-2324941.
\end{acks}

\bibliographystyle{ACM-Reference-Format}

\appendix
\section{Stress-Adjusted Water Modeling and Optimization Formulation}
\label{app:stress_modeling}

This appendix presents the formal modeling details underlying our analysis.
We define stress-adjusted water consumption, explicitly decompose on-site and off-site impacts, and describe the offline optimization formulation used to compute an upper bound on achievable sustainability gains from workload scheduling.

\subsection{Stress-Adjusted Water Consumption}
\label{app:stress_adjusted}

Traditional volumetric water accounting implicitly assumes that all water consumption has a uniform environmental impact.
In reality, the consequences of water withdrawal depend strongly on regional availability and seasonal scarcity.
To capture this heterogeneity, we define \emph{stress-adjusted water consumption} as volumetric water use weighted by a location- and time-specific water-stress factor.

We adopt the \textbf{A}vailable \textbf{WA}ter \textbf{RE}maining for the \textbf{U}nited \textbf{S}tates (AWARE-US) model, which provides county-level monthly water stress characterization.
Higher AWARE values correspond to more water-stressed conditions.

For a data center $n$ at time $t$, the total stress-adjusted water impact is the sum of on-site and off-site components:
\begin{equation}
w^{\text{stress}}_{n,t} =
\sigma^{on}_{n,t} \cdot w^{on}_{n,t}
+ \sigma^{off}_{n,t} \cdot w^{off}_{n,t},
\end{equation}
where $\sigma^{on}_{n,t}$ denotes the local water stress factor at the data center location, and $\sigma^{off}_{n,t}$ captures the water stress associated with electricity generation supplying the data center.

\subsection{On-Site Water Consumption}
\label{app:onsite}

On-site water consumption arises primarily from evaporative cooling systems, such as cooling towers.
We model on-site water use as proportional to IT power consumption:
\begin{equation}
w^{on}_{n,t} = \omega^{on}_{n,t} \cdot p_{n,t},
\end{equation}
where $\omega^{on}_{n,t}$ (L/kWh) represents the on-site water usage effectiveness under local weather and cooling conditions, and $p_{n,t}$ is the IT power consumption at data center $n$ and time $t$.
The on-site stress factor $\sigma^{on}_{n,t}$ is derived directly from the AWARE-US score of the county in which the data center is located.

\subsection{Off-Site Water Consumption from Electricity Generation}
\label{app:offsite}

In addition to on-site water use, data centers incur indirect (off-site) water consumption through electricity generation.
Many electricity generation technologies, including thermal and nuclear plants, consume water for cooling.
We model off-site water consumption as:
\begin{equation}
w^{off}_{n,t} = \omega^{off}_{n,t} \cdot \eta_n \cdot p_{n,t},
\end{equation}
where $\eta_n$ is the power usage effectiveness (PUE) of data center $n$, and $\omega^{off}_{n,t}$ denotes the average water intensity of electricity supplied to the data center at time $t$.

Crucially, electricity consumed by a data center may originate from multiple power plants located in different regions.
To account for this geographic decoupling, we compute $\sigma^{off}_{n,t}$ as the generation-weighted average water stress of upstream power plants.
This provenance-aware modeling avoids the systematic bias introduced by region-level or national-average abstractions.

\subsection{Carbon Emissions}
\label{app:carbon}

Carbon emissions are treated separately because their environmental impact is global rather than location-specific.
We compute carbon emissions as:
\begin{equation}
c_{n,t} = \gamma_{n,t} \cdot \eta_n \cdot p_{n,t},
\end{equation}
where $\gamma_{n,t}$ denotes the carbon intensity of electricity (e.g., kg CO$_2$/kWh) associated with the grid supplying data center $n$ at time $t$.

\subsection{Offline Stress-Adjusted Water Scheduling Formulation}
\label{app:offline_opt}

To quantify the maximum achievable benefit of workload shifting, we formulate an offline optimization problem that assumes perfect future knowledge of water stress, water efficiency, and carbon intensity signals.
This formulation is used solely for evaluation and benchmarking and is not intended as an online scheduling algorithm.

We consider a time-slotted horizon $t \in \{1,\ldots,T\}$, $G$ workload gateways indexed by $g$, and $N$ data centers indexed by $n$.
At each time $t$, the workload arriving at gateway $g$ is denoted by $\lambda_{g,t}$.
Workload may be executed immediately or deferred by up to $K$ time slots.

Let $x_{g,n,t,k}$ represent the amount of IT power (or power-equivalent workload) originating from gateway $g$ at time $t$ that is scheduled to execute at data center $n$ after a delay of $k$ slots.
Performance overheads, such as routing latency or migration costs, are modeled using a multiplicative factor $h_{g,n}$.

The total IT power at data center $n$ and time $t$ is:
\begin{equation}
p_{n,t} = \sum_{g=1}^{G}\sum_{k=0}^{K} x_{g,n,t-k,k} \cdot h_{g,n}.
\end{equation}

The objective minimizes a weighted combination of total carbon emissions and total stress-adjusted water impact:
\begin{equation}
\min_{\{x_{g,n,t,k}\}}
\sum_{t=1}^{T}\sum_{n=1}^{N}
\left(
c_{n,t}
+
\alpha \cdot
\left(
\sigma^{on}_{n,t} w^{on}_{n,t} + \sigma^{off}_{n,t} w^{off}_{n,t}
\right)
\right),
\end{equation}
where $\alpha$ controls the trade-off between water and carbon.

The optimization is subject to the following constraints:
\begin{align}
p_{n,t} &\le P_n, \quad \forall n,t, \\
\sum_{n=1}^{N}\sum_{k=0}^{K} x_{g,n,t,k} &= \lambda_{g,t}, \quad \forall g,t, \\
x_{g,n,t,k} &\ge 0, \quad \forall g,n,t,k .
\end{align}

By varying $\alpha$, the formulation traces the Pareto frontier between carbon- and water-centric operating regimes.

\end{document}